\documentclass[aps,prd,onecolumn,superscriptaddress,nofootinbib, a4paper%,fleqn%, showpacs
]{revtex4}
\usepackage[all]{xy}
\usepackage{graphicx}
\usepackage{slashed}
\usepackage{amsmath,bbm,latexsym,amssymb}
\usepackage{epsfig}
\usepackage{epstopdf}
\usepackage{float} 
\usepackage{tensor}
\usepackage{booktabs}
\usepackage{multirow}
\usepackage{cancel}

\def\ifmath#1{\relax\ifmmode #1\else $#1$\fi}

\renewcommand{\Re}{{\rm Re}}
\renewcommand{\Im}{{\rm Im}}

\newcommand{\be}{\begin{equation}}
\newcommand{\ee}{\end{equation}}
\newcommand{\ba}{\begin{eqnarray}}
\newcommand{\ea}{\end{eqnarray}}

\newcommand{\rep}[1]{\ensuremath\boldsymbol{#1}}

\newcommand{\Z}[1]{\ensuremath{\mathbbm{Z}_{#1}}} % z_N ->\Z{N}
\newcommand{\SO}[1]{\ensuremath{\mathrm{SO}(#1)}}
\newcommand{\SU}[1]{\ensuremath{\mathrm{SU}(#1)}}

\newcommand{\U}[1]{\ensuremath{\mathrm{U}(#1)}}

\newcommand{\Inv}[1]{\ensuremath{\mathcal{I}_{{#1}}}}
\newcommand{\Jnv}[1]{\ensuremath{\mathcal{J}_{{#1}}}}

\usepackage{xparse}

\usepackage[hyperfootnotes=false]{hyperref}
\hypersetup{
    colorlinks=true,
    linkcolor=blue,
    filecolor=blue,      
    urlcolor=blue,
    citecolor=blue
}

\makeatletter
\renewcommand*\env@matrix[1][\arraystretch]{%
  \edef\arraystretch{#1}%
  \hskip -\arraycolsep
  \let\@ifnextchar\new@ifnextchar
  \array{*\c@MaxMatrixCols c}}
\makeatother

\usepackage{tabularx}
\newcolumntype{C}[1]{>{\centering\arraybackslash}p{#1}}

\usepackage{xcolor}

\definecolor{darkgreen}{HTML}{109930}

\begin{document}
\vspace*{1cm}

\title{A fully basis invariant Symmetry Map of the 2HDM}

\author{Miguel P.\ Bento}%
\email{miguel.pedra.bento@tecnico.ulisboa.pt}
\affiliation{CFTP, Departamento de F\'{\i}sica, Instituto Superior T\'{e}cnico, Universidade de Lisboa, Avenida Rovisco Pais 1, 1049 Lisboa, Portugal}
\author{Rafael Boto}%
\email{rafael.boto@tecnico.ulisboa.pt}
\affiliation{CFTP, Departamento de F\'{\i}sica, Instituto Superior T\'{e}cnico, Universidade de Lisboa, Avenida Rovisco Pais 1, 1049 Lisboa, Portugal}
\author{Jo\~ao P.\ Silva}%
\email{jpsilva@cftp.ist.utl.pt}
\affiliation{CFTP, Departamento de F\'{\i}sica, Instituto Superior T\'{e}cnico, Universidade de Lisboa, Avenida Rovisco Pais 1, 1049 Lisboa, Portugal}
\author{Andreas Trautner}%
\email{trautner@mpi-hd.mpg.de}
\affiliation{Max-Planck-Institut f\"ur Kernphysik, Saupfercheckweg 1, 69117 Heidelberg, Germany}

%\date{\today}

\begin{abstract}
We derive necessary and sufficient conditions for all global symmetries of the most general two Higgs doublet model (2HDM) scalar potential
entirely in terms of reparametrization independent, i.e.\ basis invariant, objects.
This culminates in what we call a ``Symmetry Map'' of the parameter space of the model
and the fundamental insight that there are, in general, two algebraically distinct ways of how symmetries manifest themselves
on basis invariant objects: either, basis invariant objects can be non-trivially related, or, basis covariant objects can vanish.
These two options have different consequences on the resulting structure of the ring of basis invariants and on the number of remaining physical parameters.
Alongside, we derive for the first time necessary and sufficient conditions for CP conservation in the 2HDM entirely in terms of CP-even quantities.
This study lays the methodological foundation for analogous investigations of global symmetries in all other models that have unphysical freedom of reparametrization, most notably the Standard Model flavor sector.
\end{abstract}

\maketitle

\section{Introduction}
\label{sec:intro}
When a model has more than one field with the same quantum numbers,
the theory may be written in an infinite set of different bases,
each obtained by unitary transformations (to keep the kinetic terms canonical) 
among equivalent fields.
This obscures the determination of the independent degrees of freedom
(dof) of the theory.
The issue is even more complicated in the case of charge-parity (CP) odd
degrees of freedom, since simple rephasings of the fields bring
potentially CP-odd phases in and/or out of the Lagrangian.
The solution is to look for quantities which are basis invariant.
The prototypical example is the Jarlskog invariant \cite{Jarlskog:1985ht,Jarlskog:1985cw,Bernabeu:1986fc},
identifying the single CP violating phase present in the Standard Model (SM).

A general solution was described in \cite{Botella:1994cs}:
the couplings in the Lagrangian are defined as tensors in the spaces of
fields with identical quantum numbers;
multiplying tensors in any combination and taking traces
over all internal spaces,
leaves a physical basis invariant quantity. Its real part is
CP-even, its imaginary part (when it exists) is CP-odd.
Although correct, this solution has several problems.
First, there are infinitely many such invariants;
almost all being combinations of simpler invariants.
In fact, any theory has only a finite number of dof; 
how is one to determine it?
Relatedly,
when should one ``stop looking'' for further independent basis invariants?
Second, what are the necessary and sufficient
conditions for some property (say, CP conservation)?
Most often, one condition (necessary or sufficient)
is much simpler to address than the other.
For decades, these issues were addressed on a model-by-model basis.

Around 2007, the issue of basis invariants in particle physics
was put on a sound mathematical ground.
Basis invariants were seen to form a ring, in the algebraic sense,
and techniques involving the Hilbert-Poincar\'{e} series (HS)
and the Plethystic logarithm (PL) were imported into
particle physics by Hanany and collaborators
\cite{Benvenuti:2006qr,Feng:2007ur}.
It has been applied to a number of formal questions \cite{Noma:2006pe,Butti:2007jv,Gray:2008yu,Hanany:2008kn,Hanany:2008sb,Hanany:2014dia,Bourget:2017tmt},
as well as the determination of quark and lepton
invariants in and beyond the SM \cite{Jenkins:2007ip,Jenkins:2009dy,Hanany:2010vu}.
Recently \cite{Trautner:2018ipq}, the HS and PL techniques were used in order to
determine the number of independent basis invariants, a generating set
of basis invariants, and the structure of relations between basis
invariants (the so-called syzygies) in the most general two Higgs doublet model (2HDM).

Given some theory and quantum fields,
one can impose additional discrete or continuous symmetries
on the fields,
either family symmetries, relating fields with equal quantum numbers,
or so-called generalized CP symmetries (GCP), relating fields
with linear combinations of the CP transformed fields.
In the context of theories with N Higgs doublets (NHDM),
these are known as symmetry-constrained models.
A well known example is the two Higgs doublet model,
where a \Z2 symmetry ($\phi_1 \rightarrow \phi_1$,
$\phi_2 \rightarrow - \phi_2$) is imposed.
Given the limited number of fields and the limited number of
Lagrangian terms, many imposed symmetries may yield the same
symmetry-constrained model.
That is, two sets of apparently different symmetries may yield
models which have the same number of dof and make exactly
the same physical predictions.\footnote{Sometimes this equality may be obscured
because the symmetries are written in different bases.}
For example,
imposing \Z3 and \Z4 on the scalar sector of the 2HDM
gives exactly the same potential \cite{Ferreira:2008zy}.
In this context,
two further interesting questions arise:
\begin{enumerate}
\item What types of physically distinct symmetry-constrained models
can one obtain by applying family and/or GCP symmetries in a given theory?
How many independent possibilities are there?
\item What will be the relations between invariants in the general theory
that define a given symmetry-constrained model?
\end{enumerate}
There is no known solution or general technique for either of these questions.

However, some work has been done in specific cases.
For example,
regarding the first question,
it has been proved that
there are only six symmetry-constrained 2HDM models \cite{Ivanov:2006yq},
dubbed in \cite{Ferreira:2009wh,Ferreira:2010yh} as \Z2, \U1,
\SO3, CP1, CP2, and CP3 (for a review, see \cite{Branco:2011iw}).
The classification has also been achieved in the 3HDM \cite{Ferreira:2008zy,Ivanov:2011ae,Ivanov:2012ry,Ivanov:2012fp},
but results in NHDM for $N>3$ are unknown.
The second question has also been addressed in the context of the 2HDM,
but always using the tensorial technique or a related bilinear space technique.
Examples include
\cite{Davidson:2005cw,Ivanov:2006yq,Ferreira:2010yh}.
Neither question has been tackled using techniques based on rings of invariants.
We suspect that both questions will ultimately be simpler to solve in
this framework.

In this article,
we use the basis invariants found in \cite{Trautner:2018ipq}
for the most general 2HDM,
in order to tackle the second question,
obtaining a complete map of invariant relations and how they give rise to
augmented symmetries.
We will re-obtain some results found previously,
but we will also obtain new results, even in this simple context.
Most importantly,
we will show how these features are related with the presence of
several sub-rings of invariants.
Many results which would seem ad-hoc before get here their full
interpretation.

One interesting instance concerns the relations needed to increase the symmetry
of the model versus the number of dof lost in the process.
For example,
the most general 2HDM has eleven independent dof,
while the CP conserving CP1 has nine\footnote{Table 5 of \cite{Branco:2011iw}
has a misprint on its last line. It states 8 dof when it should state 9 dof
for the CP1 case.};
thus, two dof are lost when the symmetry is increased.
Nevertheless,
as shown in \cite{Gunion:2005ja},
four equations are needed in order to define 
the CP1 model.
The difference between the four equations needed to define the
increase in symmetry and the mere two dof reduced in the process
is described in the usual analysis as a consequence of the
existence of special regions in parameter space.
Finding those regions always seems a hap-hazard process.
We will show that this has a very simple explanation in algebraic
terms, as there exist in fact several rings of invariants in the 2HDM.
We will show how this explains the connection
between relations needed and number of dof lost,
for any symmetry increase one might wish to consider.

We introduce the tensor and the bilinear notations for the 2HDM
in section~\ref{sec:one},
where we also discuss the global symmetries and the degenerate regions of
parameter space.
In section~\ref{sec:two},
we use for the first time the invariant formalism of \cite{Trautner:2018ipq}
in order to fully describe all symmetry-constrained 2HDM models.
We show that there are multiple rings of invariants and construct the
``Symmetry Map'' for the 2HDM,
shown in Figure~\ref{SymmetryMap},
pointing out that there are two algebraically different ways
to move along this map and that they have different consequences.
We explain how this solves the relations needed versus number
of dof lost puzzle.
In the process,
we find some other new results.
For example,
in section~\ref{sec:CP1primary} we prove 
that the necessary and sufficient conditions for CP conservation
can be expressed solely in terms of CP-even invariants.
This possibility in known in the SM but was so far unknown in the 2HDM.
Further,
we show in section~\ref{sec:z2} that there are three equations
necessary and sufficient to ascend from the CP1 model to 
the \Z2 model, although only two dof are lost in the process. 
Our main results are compiled
in Figure~\ref{SymmetryMap}
and Tables~\ref{tab:invariants}-\ref{tab:summary},
and summarized in section~\ref{sec:summary}.
We present our conclusions in section~\ref{sec:conclusions}.
The appendices contain material that would distract from the flow of the text,
namely, connections with other notations,
detailed calculations needed in the text,
and a discussion of the syzygies for the 2HDM invariant ring,
including several new results.

\section{The 2HDM scalar sector}
\label{sec:one}
\subsection{Different notations for the scalar potential}
The scalar sector of the 2HDM consists of two hypercharge-one $\SU2_\mathrm{L}$  doublets $\Phi_a(x)\equiv (\phi^+_a(x)\,,\,\phi^0_a(x))^{\mathrm{T}}$ which we 
label by a ``Higgs flavor'' index $a=1,2$.
In the conventional parametrization the most general renormalizable $\SU2_\mathrm{L}\times \U1_\mathrm{Y}$ invariant scalar potential is given by
\begin{equation}\label{potV}
\begin{split}
 V &= m_{11}^2\,\Phi_1^\dagger\Phi_1 + m_{22}^2\,\Phi_2^\dagger\Phi_2 - \left[m_{12}^2\,\Phi_1^\dagger\Phi_2 + {\rm h.c.}\right]\\
   &\quad + \frac{\lambda_1}{2}\left(\Phi_1^\dagger\Phi_1\right)^2 + \frac{\lambda_2}{2}\left(\Phi_2^\dagger\Phi_2\right)^2 + \lambda_3\left(\Phi_1^\dagger\Phi_1\right)\left(\Phi_2^\dagger\Phi_2\right) + \lambda_4\left(\Phi_1^\dagger\Phi_2\right)\left(\Phi_2^\dagger\Phi_1\right)\\
   &\quad + \left\{\frac{\lambda_5}{2}\left(\Phi_1^\dagger\Phi_2\right)^2 + \left[\lambda_6\left(\Phi_1^\dagger\Phi_1\right) + \lambda_7\left(\Phi_2^\dagger\Phi_2\right)\right]\Phi_1^\dagger\Phi_2 + {\rm h.c.} \right\}\,.  
\end{split}
\end{equation}
Here $m_{11}^2$, $m_{22}^2$, and $\lambda_{1,2,3,4}$ are real parameters, while $m_{12}^2$, $\lambda_5$, $\lambda_6$, and $\lambda_7$ are in general complex.
Not all of these fourteen real parameters are physical due to the \textit{unphysical} freedom to change the Higgs flavor basis by a 
unitary transformation
\be\label{Basischange} 
\Phi_a'=U_{ab}\Phi_b, \quad U \in \U2\;.
\ee
Here and in the following the summation over repeated indices is implicit. 
It happens that the $\U1$ factor in $\U2\cong\SU2\times\U1$ coincides with the $\U1_\mathrm{Y}$ hypercharge transformation, 
which is a gauged symmetry. Therefore, the global rephasing of the Higgs fields does not have any effect 
on the Higgs potential implying that we do not have to take the \U1 factor into account in all subsequent considerations.
In order to assign a meaning to the parameters appearing in \eqref{potV}
it is required to make a \textit{basis choice} for the scalar fields. 
Utilizing all possible basis changes to absorb parameters one can show that 
the actual number of real physical parameters is $11=14-3$, where three is the number
of generators of the \SU2 basis rotation \cite{Santamaria:1993ah}. 

An alternative, \SU2 covariant notation for the scalar potential, following \cite{Trautner:2018ipq} and introduced in \cite{Botella:1994cs,Gunion:2005ja}, is  
\be \label{genericpot}
V=\Phi_a^\dagger \tensor{Y}{^a_b} \Phi^b
+\Phi_a^\dagger\Phi_b^\dagger\tensor{Z}{^a^b_c_d}\Phi^c\Phi^d,\quad 
\ee
where $a, b, c, d=1, 2$ are indices in Higgs flavor space. 
Here we introduce a notation with upper and lower indices to distinguish fields transforming as $\rep{2}$ or $\rep{\bar{2}}$ under Higgs basis changes. 
Hermiticity of $V$ and $\SU2_{\mathrm{L}}$ gauge invariance imply that 
\be
\tensor{Y}{^a_b}=\left(\tensor{Y}{^b_a}\right)^*, \quad \tensor{Z}{^a^b_c_d}=\left(\tensor{Z}{^c^d_a_b}\right)^*,\quad\text{as well as}\quad \tensor{Z}{^a^b_c_d}=\tensor{Z}{^b^a_d_c}\;.
\ee
Under the global \SU2 basis change \eqref{Basischange}, the tensors $Y$ and $Z$ transform as
\begin{align}
\tensor{Y}{^{\prime a}_b}&=\tensor{U}{^a_{a'}}\,\tensor{Y}{^{a'}_{b'}}\,\tensor{U}{^{\dagger b'}_b}\;, \\
\tensor{Z}{^{\prime a}^b_c_d}&=\tensor{U}{^a_{a'}}\tensor{U}{^b_{b'}}\,\tensor{Z}{^{a'}^{b'}_{c'}_{d'}}\,\tensor{U}{^{\dagger c'}_c}\tensor{U}{^{\dagger d'}_d}\;,
\end{align}
where we now also use the matrix $U$ with upper and lower indices according to our conventions.

\subsection{Basis invariant quantities}
\label{sec:BIs}
From the basis covariant quantities $Y$ and $Z$ it is, in principle, straightforward to obtain basis invariant quantities simply by a complete contraction of indices \cite{Botella:1994cs,Gunion:2005ja,Davidson:2005cw}.
A more systematic method for the explicit construction of basis invariants has been presented in \cite{Trautner:2018ipq}.
There, basis invariant quantities are classified depending on their content of basis covariant quantities. Let us briefly summarize these results.

First, one finds \textit{linear} combinations of the entries of the tensors $Y$ and $Z$ (corresponding to linear combinations of the $m^2$'s and $\lambda$'s of \eqref{potV}) 
which transform in \textit{irreducible} representations of the \SU2 group of basis changes. These form the so-called \textit{building blocks} for all further considerations, 
including the construction of \textit{non-linear} higher-order basis invariants.

For the 2HDM one finds that there are three algebraically independent linear combinations of potential parameters which are basis invariant already by themselves.
Those are given by\footnote{%
We will relate our invariants and invariant relations to earlier works in the literature (see~\cite{Ivanov:2005hg,Nishi:2006tg,Maniatis:2007vn,Ivanov:2006yq},
and especially \cite{Branco:2011iw} and references therein) mostly following the notation of Nishi~\cite{Nishi:2006tg}.
The singlets can be written as linear combinations of the singlets in~\cite{Nishi:2006tg} as 
$Y_{\rep{1}}=\mathrm{M}_0$, $Z_{\rep{1}_{(1)}}=\frac{1}{4}(3\,\Lambda_{00}+\,\mathrm{tr}\tilde{\Lambda})$, and $Z_{\rep{1}_{(2)}}=\frac{1}{4}(\Lambda_{00}-\mathrm{tr}\tilde{\Lambda})$.}
\be\label{singlets}
 Y_{\rep{1}}:=\tensor{Y}{^a_a},\quad Z_{\rep{1}_{(1)}}:=\frac12\left(\tensor{Z}{^{ab}_{ab}}+\tensor{Z}{^{ab}_{ba}}\right),\quad \text{and}\quad Z_{\rep{1}_{(2)}}:=\varepsilon_{ab}\,\varepsilon^{cd}\,\tensor{Z}{^{ab}_{cd}}\;,
\ee
where $\varepsilon$ is the total anti-symmetric tensor in the convention $\varepsilon^{12}=1$.
Further, one finds three covariantly transforming building blocks denoted by
\be\label{multiplets}
 Y_{\rep{3}}\equiv \mathrm{Y},\quad Z_{\rep{3}}\equiv\mathrm{T},\quad \text{and} \quad Z_{\rep{5}}\equiv\mathrm{Q}\;.
\ee
These transform in the triplet ($\mathrm{Y}$ and $\mathrm{T}$) and quintuplet ($\mathrm{Q}$) representation under \SU2 basis changes,
and we have also introduced the shorthand notation ($\mathrm{Q}$, $\mathrm{Y}$, $\mathrm{T}$) for them for later use.
For general explicit expressions for these, as well as for explicit expressions for all building blocks in terms of the parametrization of \eqref{potV}
we refer the reader to Ref.\ \cite[Eqs.(28,63)]{Trautner:2018ipq}.
To fully characterize the ring of basis invariants, including their structure, 
the (multi-graded) Hilbert Series together with the Plethystic logarithm are used (see e.g.\ \cite{Benvenuti:2006qr,Feng:2007ur} and \cite{Lehman:2015via} for an introduction).
For completeness we state the HS for the most general 2HDM scalar potential in Appendix~\ref{app:QYTring}.
This procedure informs us that the smallest complete set of algebraically independent invariants contains four invariants of order $2$ (in the building blocks), 
three of order $3$, and one invariant of order $4$. We follow \cite{Trautner:2018ipq} and denote invariants by 
\be\label{eq:invariants}
\Inv{a,b,c}\equiv\left[\mathrm{Q}^a\mathrm{Y}^b\mathrm{T}^c\right]\quad\text{for invariants that contain powers} 
\quad Z_{\rep{5}}^{\otimes a}\otimes Y_{\rep{3}}^{\otimes b}\otimes Z_{\rep{3}}^{\otimes c}
\ee
of the building blocks. 
The set of algebraically independent invariants chosen in \cite{Trautner:2018ipq} is
\ba\label{setIs}
\mathcal{I}_{2,0,0},\quad \mathcal{I}_{0,2,0},\quad \mathcal{I}_{0,0,2},\quad \mathcal{I}_{0,1,1},\quad \mathcal{I}_{3,0,0},
\quad \mathcal{I}_{1,2,0},\quad \mathcal{I}_{1,0,2},\quad\text{and}\quad\mathcal{I}_{2,1,1}\;.
\ea
Basis invariants in this convention are CP-even(odd) if and only if they contain an even(odd) number of triplet building blocks~\cite{Trautner:2018ipq}.
Hence, all of the above invariants are CP-even.
Beyond this set of algebraically independent invariants, 
there is the so-called generating set of a ring of invariants, 
consisting of invariants that cannot be written as a polynomial of other invariants. 
In the 2HDM, this set contains eleven additional invariants (with CP-odd invariants denoted by $\mathcal{J}$ instead of $\mathcal{I}$),
\ba\label{generatingSet}
\mathcal{I}_{1,1,1},\quad \mathcal{I}_{2,2,0},\quad \mathcal{I}_{2,0,2},\quad \mathcal{J}_{1,2,1},\quad \mathcal{J}_{1,1,2},
\quad \mathcal{J}_{2,2,1},\quad \mathcal{J}_{2,1,2},\quad \mathcal{J}_{3,3,0},\quad \mathcal{J}_{3,0,3},\quad \mathcal{J}_{3,2,1},\quad\text{and}\quad\mathcal{J}_{3,1,2}\;.
\ea
The explicit form of all invariants has been obtained by the use of Young tableaux and the corresponding hermitian projector operators 
and has been given in \cite{Trautner:2018ipq}.
For convenience of the reader we also state some of the invariants in the conventional parametrization of the potential and the common special choice of basis where 
$\lambda_6=-\lambda_7$ in Appendix~\ref{app:InvariantsInConvPar}.

Several comments are in order. Note that the (total dimension of) building blocks stated in eqs.\ \eqref{singlets} and \eqref{multiplets} 
together reflect the fourteen real degrees of freedom of the generic potential \eqref{potV}. By contrast,
the total number of algebraically independent basis invariants from eqs.\ \eqref{singlets} and \eqref{setIs} is eleven,
corresponding to the total number of eleven independent physical parameters of the 2HDM scalar sector.
Alternative methods for the construction of basis invariants in the 2HDM exist, see e.g.\ \cite{Gunion:2005ja,Davidson:2005cw}
and \cite{Ivanov:2005hg,Nishi:2006tg,Ivanov:2006yq,Bednyakov:2018cmx}. We will continuously point out the overlap of our derived relations
with these approaches in a set of footnotes.

\subsection{Global symmetries of the scalar potential}
\label{sec:Symmetries}
Let us now discuss exact global symmetries of the Higgs potential. 
As summarized in \cite{Branco:2011iw} these symmetries can be classified into two types:
\begin{itemize}
    \item Higgs flavor (HF) symmetries,
    \be\label{HFsym}
    \Phi^a\rightarrow\left(\Phi^S\right)^a:=\tensor{S}{^a_b}\Phi^b\;,
    \ee
    with a unitary matrix $S$. 
    \item General CP (GCP) symmetries,
    \be\label{GCPsym}
    \Phi^a\rightarrow\left(\Phi^{GCP}\right)_a:=
    \Phi_b^*\tensor{\left[X^\mathrm{T}\right]}{^b_a}\;, 
    \ee
    with a unitary matrix $X$.
\end{itemize}
Under a basis change of the form \eqref{Basischange}, the HF and GCP transformation matrices transform accordingly,
\begin{align}
S' &= U\,S\,U^\dagger, \\
X' &= U\,X\,U^\mathrm{T}.
\end{align}
Assuming invariance of the potential under a HF or GCP transformation requires relations among the potential parameters. 
In general, this reduces the number of independent parameters and, likewise, also the number of independent basis invariants.
While many different symmetries are thinkable, some of them actually enforce parameter relations which give rise to
a higher, ``accidental'' symmetry than initially required (this is an effect of taking into account only gauge invariant operators up to mass dimension four 
in the scalar potential for the sake of renormalizability). 
We focus on symmetries which do not give rise to accidental symmetries. Those are 
called ``realizable''.
Ivanov has shown that there are only six distinct classes of potentials corresponding to six 
distinct realizable symmetries \cite{Ivanov:2006yq}. 
These are commonly denoted as $\Z2$, \U1, and \SU2 as well as CP1, CP2, CP3, and they are schematically related as (see \cite{Ferreira:2009wh,Ferreira:2010yh,Branco:2011iw} for extensive discussions)
\begin{equation}\label{symtree}
 \mathrm{CP1}~\subset~\Z2~\subset~\left\{\begin{array}{cc} \U1 \\ \mathrm{CP2} \end{array}\right\}~\subset~\mathrm{CP3}~\subset~\SU2\;.
\end{equation}
We will focus on those six realizable symmetries here and analyze them in terms of basis invariants and their interrelations.
While the explicit form of the transformations \eqref{HFsym} and \eqref{GCPsym} clearly depend on the basis chosen for the Higgs doublets,
the relations we will identify among the basis invariants are inherently basis independent.

\subsection{``Degenerate regions'' of parameter space and ``Symmetry Map'' of the 2HDM}
\label{sec:degeneracies}
Before we start the discussion of basis invariant relations, we make the following important observation: 
The structure of the 2HDM ring is only as discussed in section \ref{sec:BIs} if indeed \textit{all} of the non-trivial
building blocks, Q, Y, and T are non-vanishing, and covariant building blocks transforming in the same irreducible representation (here Y and T) are not \textit{aligned}.\footnote{%
For the present case of the 2HDM the \textit{alignment} of building blocks Y and T literally corresponds the alignment of the two three-dimensional 
vectors $\Vec{\text{M}}$ and  $\Vec{\Lambda}$ in the geometric language~\cite{Nishi:2006tg}. 
However, we stress that for more general problems (in particular for higher dimensional representations and freedom of basis-changes beyond \SU2) 
\textit{alignment} of two building blocks can still be algebraically formulated but does not necessarily always have to correspond to any particularly nice geometrical intuition.}
If any of the building blocks vanishes, or if identically transforming covariants Y and T are aligned, then the ring changes its structure and, in principle,
a different (smaller) ring should be discussed. Regions in the parameter space where this happens have previously been called ``special'' \cite{Gunion:2005ja} 
or ``degenerate'' \cite{Ivanov:2005hg}. Here, these degenerate regions are understood to be special, in the sense that they lead to rings with a different ``topology'' than the 
original ring.
In a fully basis invariant language, the special regions in parameter space are given by 
\begin{align}\label{eq:DegCases}
 \mathrm{(I)} \quad \mathrm{Q} &= 0\;,& 
 \mathrm{(II)} \quad \mathrm{Y} &= 0\;,&
 \mathrm{(III)} \quad \mathrm{T} &= 0\;,&
 \mathrm{(IV)} \quad \left(\mathrm{YT}\right)^2 &= \mathrm{Y}^2\mathrm{T}^2\;.&
\end{align}
Regions (I)-(III) imply the vanishing of all invariants that contain the corresponding building block.
Region (IV) effectively turns out to be very similar to regions (II) and (III), as 
there will be only one independent direction of a triplet building block which we will discuss in detail below.
In fact, regions (II) and (III) turn out to be identical from an invariant theory point of view,
in the sense that they give rise to identical rings. 
However, in the full quantum theory, when renormalization group running is taken into account,
these regions can be distinguished as we will discuss below in section~\ref{sec:InvsandRGE}.
All of the above regions give rise to very different rings than the full 2HDM ring discussed in section~\ref{sec:BIs}.
All of these rings are smaller than the full 2HDM ring, and can be obtained from it by ``collapsing'' the ring, i.e.~setting to zero the corresponding variables.

We will see that certain symmetries enforce one or several of the degenerate regions above.
That is, certain symmetries cannot be realized if certain building blocks are non-vanishing.
This is very important because it implies that there are, in general, two different ways how to move in the ``space''
of potential symmetries: 
\begin{enumerate}
 \item One can impose relations amongst certain (primary) basis invariants, or
 \item One can impose the vanishing of certain building blocks of basis invariants.
\end{enumerate}
While the first possibility operates within a given ring and leaves the 
ring ``intact'', the second possibility ``collapses'' the ring to a (potentially much) smaller ring, and the discussion of further
symmetries then must be based on this smaller ring. To further highlight the important distinction between options 1. and 2. above, 
we will see below that the first option generally reduces the number of parameters in one-to-one correspondence with the number of imposed relations. 
By contrast, for the second option the number of independent parameters can be vastly reduced by imposing the vanishing of a single building block.
All this will be explored in detail below, Nonetheless let us give an example: 
Only if all building blocks are non-vanishing and only if there is no alignment of Y and T,
then the full number of $11$ independent parameters of the 2HDM ring can be supported. 
In contrast, requiring region (IV) above, i.e.~alignment of Y and T -- a single relation --, 
the ring to be discussed (the corresponding ring is discussed in detail in Appendices~\ref{app:YTRing} and~\ref{app:YTRingAl}) 
turns out to only have $9$ independent parameters.\footnote{%
That removing one parameter from a model can have an effect on other physical parameters is not an unusual behavior. 
Think, for example, about the SM where nulling a quark mixing angle also removes the physicality of the CP-odd phase;
or a recent example \cite{Cherchiglia:2020kut}, where removing a CP-odd physical quantity also removes a CP-even physical parameter.
However, the origin of such incidences has, to our knowledge, never been discussed from a basis invariant viewpoint.}

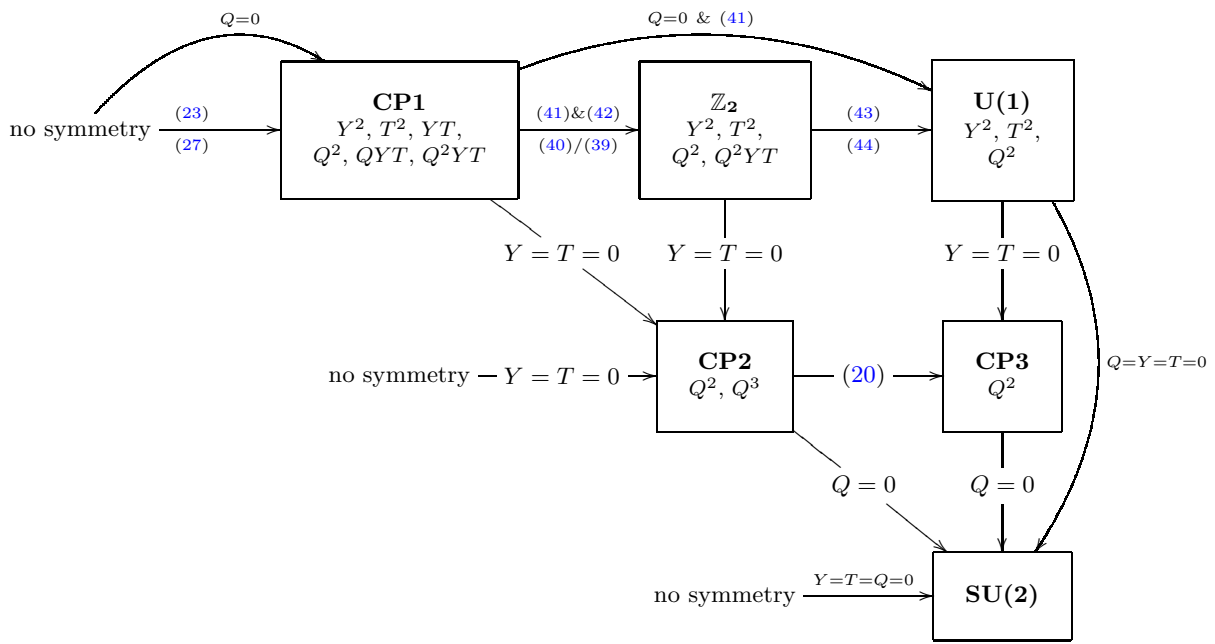
\begin{figure}[t]\renewcommand{\arraystretch}{1.5}
\hspace{\fill}%
 \xymatrix@=1.6cm{
 \text{no symmetry} \ar[r]^(.35){\eqref{CPCnodeg}}_(.35){\eqref{eq:CPCdeg}} \ar@/^3pc/[r]^{Q=0} &
 *++++[F]{\txt{\textbf{CP1}\\ \footnotesize{$Y^2$, $T^2$, $YT$,} \\ \footnotesize{$Q^2$, $QYT$, $Q^2YT$}} } 
          \ar[r]^(.55){\eqref{eq:Z2022}\&\eqref{eq:Z2222}}_(.55){\eqref{Z2:R204}/\eqref{Z2:R240}} \ar[dr]|*+{Y=T=0} \ar@/^3pc/[rr]^{Q=0~\&~\eqref{eq:Z2022}}  &  
 *++++[F]{\txt{$\boldsymbol{\mathbb{Z}_{2}}$ \\ \footnotesize{$Y^2$, $T^2$,} \\ \footnotesize{$Q^2$, $Q^2YT$}} }   \ar[r]^{\eqref{Z2toU1A}}_{\eqref{U1:I202b}} \ar[d]|*+{Y=T=0} & 
 *++++[F]{\txt{\textbf{U(1)}\\ \footnotesize{$Y^2$, $T^2$,} \\ \footnotesize{$Q^2$} }} \ar[d]|*+{Y=T=0} \ar@/^3pc/[dd]^{Q=Y=T=0} \\ 
 &
 \text{no symmetry} \ar[r]|*+{Y=T=0} &
 *++++[F]{\txt{\textbf{CP2}\\ \footnotesize{$Q^2$, $Q^3$}}} \ar[r]|*+{\eqref{CP3:relation}} \ar[dr]|*+{Q=0} &  
 *++++[F]{\txt{\textbf{CP3}\\ \footnotesize{$Q^2$}}} \ar[d]|*+{Q=0} \\
 &
 & \text{no symmetry} \ar[r]^{Y=T=Q=0} &
 *++++[F]{\textbf{SU(2)}} \\
}
\hspace{\fill}
 \caption{\label{SymmetryMap}
 The ``Symmetry Map'' of the parameter space of the unbroken 2HDM.
 We list the classes of symmetries together with our choice of primary invariants corresponding to the number of independent parameters (for the non-degenerate case only)
 and the respective steps for symmetry enhancements.
 We do not include the three trivial singlet invariants shown in Eq.\ \eqref{singlets}, which are present for all classes of symmetries.
 All horizontal steps are given by relating previously independent, different basis invariants, while all the vertical steps are given by eliminating covariant building blocks from the ring, 
 i.e.~setting them to zero. In this sense, each horizontal line represents a ``strand'' of symmetries of an ``intact'' ring where no degeneracies arise, while moving vertically
 requires to ``collapse'' the ring to a smaller (sub-)ring by eliminating building blocks. The equation numbers \textit{above} horizontal arrows refer to sufficient relations between invariants for the non-degenerate case, 
 while equation numbers \textit{below} the arrows refer to sufficient invariant relations for the degenerate cases (II), (III) and (IV) (see text for details).
 }
\end{figure}

We illustrate these particular features in the form of a ``symmetry map'' of the 2HDM in Figure~\ref{SymmetryMap}.
Moving horizontally in this map corresponds to imposing relations among different basis invariants
and removes parameters in one-to-one correspondence with the number of relations.
In contrast, moving vertically requires the vanishing of one or several building blocks 
which can eliminate several parameters at once. Not all connections in the symmetry map
can be understood at this point of the paper, but we will gradually uncover them below.

We will now explore the action of global symmetries in terms of the basis invariants.
This shall also make the discussion of this section much clearer and gradually fill the gaps in Fig.~\ref{SymmetryMap}.

\section{The Six classes of symmetries in a basis invariant formalism}
\label{sec:two}
After imposing symmetries on the scalar potential, the number of algebraically independent invariants 
will generally be reduced, since either new relations appear or basis invariants are forced to vanish.
A practically very useful way to determine the number of algebraically independent
invariants from within a set of invariants is to determine the rank of their corresponding Jacobi matrix (see e.g.\ \cite[App.A]{Trautner:2018ipq}).
We will in the following make frequent use of this so-called Jacobi criterion to determine the number of algebraically independent invariants.
We will explicitly state the newly found relations between the basis invariants which are implied by the enhanced symmetries; 
these are relations which are \textit{necessary} for a given symmetry. But also the opposite direction will be explored: namely, 
we will also state basis invariant relations which are \textit{sufficient} for a given symmetry to be realized. 
As one of our main results, we give fully basis invariant necessary and sufficient conditions for all realizable symmetries.

In principle we would like to start from the least symmetric case of no symmetry and then move our way up to the most symmetric cases.
However, as it turns out, the least symmetric cases are the most complicated ones, and so such a procedure is not pedagogical.
Hence, we start with the most symmetric cases and move our way down to lesser symmetric cases. 
This procedure makes sense until we reach CP2. From thereon, we will switch gears and move to the least symmetric cases and 
then move our way upwards.

\subsection{\boldmath \U2 Higgs flavor symmetry\unboldmath}
\label{sec:U2}
As already discussed above, the potential is automatically invariant under the overall \U1 factor in $\U2\cong\SU2\times\U1$
for what reason we do not consider it. 
The remaining \SU2 transformation can be parametrized as
\be\label{matrixS:U2}
    S = \begin{pmatrix} e^{-i\xi}\,\cos{\theta} & e^{-i\psi}\,\sin{\theta}  \\ -e^{i\psi}\,\sin{\theta} & e^{i\xi}\,\cos{\theta}  \end{pmatrix}\;,
\ee
where $\xi$, $\theta$ and $\psi$ are three real parameters.
Requiring that the potential is invariant under a HF transformation \eqref{HFsym} with $S$ as above 
for \textit{every} $\xi$, $\theta$ and $\psi$, implies that all components of the non-trivial 
building blocks $Y_{\rep{3}}$, $Z_{\rep{3}}$ and $Z_{\rep{5}}$ are vanishing.
Consequently, all invariants in \eqref{setIs} and \eqref{generatingSet} are vanishing identically.
The set of algebraically independent invariants is then reduced to only three, 
namely $Y_{\rep{1}}$, $Z_{\rep{1}_{(1)}}$ and $Z_{\rep{1}_{(2)}}$
which transform as trivial singlets under basis changes.
Intuitively it makes sense that no non-trivial basis covariant object may exist in the space of couplings
if the full freedom of basis transformations is required as a symmetry.

The necessary and sufficient condition for \SU2 symmetry, hence, is the vanishing of all non-trivial basis
invariants. Noteworthy, this implies that \SU2 symmetry requires \textit{all} of the degenerate parameter regions (I)-(III) of Eq.~\eqref{eq:DegCases} 
to be realized [(IV) is then trivially fulfilled].

\subsection{\label{sec:cp3}CP3 symmetry}
Let us consider a less symmetric case. The transformation which is commonly referred to as CP3 is a GCP symmetry~\eqref{GCPsym} with a transformation matrix $X$ of the form
\be\label{matrixX:CP3}
    X = \begin{pmatrix} \phantom{-}\cos{\theta} & \sin{\theta}  \\ -\sin{\theta} & \cos{\theta}  \end{pmatrix}\;,
\ee
where $\theta$ is a real angle different from the special values $k\,\pi/2$ $(k\in\Z{})$.
For a scalar potential invariant under CP3, the components of those building blocks which transform as triplets under basis changes, namely $Y_{\rep{3}}$ and $Z_{\rep{3}}$, are all null.
As a direct consequence all invariants in \eqref{setIs} and \eqref{generatingSet} which contain these building blocks are identically zero. 
These are \textit{all besides} \Inv{2,0,0} and \Inv{3,0,0}.
The set of algebraically independent invariants then consists only of the singlets, $Y_{\rep{1}}$, $Z_{\rep{1}_{(1)}}$ and $Z_{\rep{1}_{(2)}}$, 
as well as possible combinations of the quintuplet building block $Z_{\rep{5}}$.
However, not all components of $Z_{\rep{5}}$ turn out to be independent.
Using the Jacobi criterion for algebraic independence of invariants one finds that there are altogether only four independent invariants.
This suggests that only one independent invariant can be built out of $Z_{\rep{5}}$ in the CP3 case. 
And indeed, requiring CP3 symmetry one finds that the a priori independent invariants \Inv{2,0,0} and \Inv{3,0,0} are related via
\be\label{CP3:relation}
\mathcal{I}_{3,0,0}^2\, =  \, \left(\tfrac{1}{3}\,\mathcal{I}_{2,0,0}\right)^3\;.
\ee
In summary, in the CP3 case the set of algebraically invariants is reduced to four.
The necessary and sufficient condition for CP3 symmetry is the vanishing of all non-trivial basis
invariants besides \Inv{2,0,0} and \Inv{3,0,0}, which, however must be related by \eqref{CP3:relation}.\footnote{%
\label{foot:CP3}%
In the language of~\cite{Nishi:2006tg} this means that the vectors $\Vec{\text{M}}$ and  $\Vec{\Lambda}$ have to vanish 
while the tensor $\tilde{\Lambda}$ is required to have two degenerate eigenvalues. The latter condition has been written in a basis invariant way as 
the vanishing of the basis invariant ``$D$'' introduced in~\cite{Ferreira:2009wh}. The vanishing of $D$ exactly corresponds to \eqref{CP3:relation} 
up to an overall numerical factor.}

\subsection{CP2 symmetry}
\label{sec:cp2}
CP2 is a GCP symmetry \eqref{GCPsym} that can be represented by the matrix \eqref{matrixX:CP3} for the special choice $\theta=\pi/2$, that is
\be\label{matrixX:CP2}
    X = \begin{pmatrix} 0 & 1 \\ -1 & 0  \end{pmatrix}\;.
\ee
CP2 again forces the triplet building blocks $Y_{\rep{3}}$ and $Z_{\rep{3}}$ to vanish. 
However, this time no relation between the components of $Z_{\rep5}$ is implied.
In agreement with that, the Jacobi criterion indicates that there are five independent invariants: the three trivial invariants plus $\mathcal{I}_{2,0,0}$ and $\mathcal{I}_{3,0,0}$.
That is, the relation \eqref{CP3:relation} is broken and no other relation of this type exists. 
We conclude that the only difference between CP2 and CP3 is the (non-)fulfillment of relation \eqref{CP3:relation}.
This also implies that one can ascend from CP2 to CP3 by enforcing \eqref{CP3:relation}.
We have checked explicitly that fulfilling \eqref{CP3:relation} (on top of a CP2 symmetry, and for $Q\neq0$) is necessary and sufficient
for increasing the symmetry to CP3.
The necessary and sufficient condition for CP2 symmetry, hence, is the vanishing of all non-trivial basis
invariants besides \Inv{2,0,0} and \Inv{3,0,0}.\footnote{%
Of course, one should also require that \eqref{CP3:relation} is \textit{not} fulfilled, otherwise the symmetry would be CP3. 
This caveat of potential higher symmetries exists for all of our necessary and sufficient conditions for symmetries, but we will
never again explicitly mention it.
It is always straightforward to check that no higher symmetry is conserved by checking the necessary and sufficient conditions for the next higher symmetry.
}

\medskip 

Together, this discussion of \SU2, CP3, and CP2 covers the lower right corner of the 2HDM symmetry map shown in Fig.~\ref{SymmetryMap}.
To elucidate the other connections we will see that it makes sense to start from the lowest symmetry, CP1, and move our way up to \U1 and the other symmetries. 

\subsection{\label{CP1}CP1 symmetry}
The smallest possible symmetry in the 2HDM is CP1 (``smallest'' in the sense that it allows for the most independent physical parameters). 
CP1 is a GCP symmetry \eqref{GCPsym} where $X=\mathbbm{1}$ can be taken to be just the identity matrix.
The prototypical example for a 2HDM invariant under this transformation is Lee's model \cite{Lee:1973iz}. 
Applying the Jacobi criterion after requiring this symmetry, one finds that the number of independent invariants
gets reduced from eleven to nine. This matches the number of physical parameters counted in \cite[p.\ 84]{Branco:2011iw}. 

A well-known straightforward basis invariant test of a realized CP1 symmetry is to check 
the vanishing of four specific CP-odd basis invariants \cite{Gunion:2005ja,Ivanov:2005hg,Nishi:2006tg,Branco:2005em,Maniatis:2007vn}, 
while making sure that no other symmetry is preserved.
In our language, the necessary and sufficient conditions for CP conservation consist of the vanishing of the four invariants
\be\label{CPC}
\Jnv{1,2,1}=\Jnv{1,1,2}=0\,,\qquad \Jnv{3,3,0}=\Jnv{3,0,3}=0\;.
\ee
A concise general proof of this, based on the interrelations of basis invariants (syzygies), was given in \cite{Trautner:2018ipq}.

As the number of independent invariants and parameters in the CP1 case is reduced only by two, one may wonder why the necessary and sufficient 
conditions for CP1 consists of four instead of two relations.
This has been understood previously \cite{Gunion:2005ja,Ivanov:2005hg,Nishi:2006tg,Maniatis:2007vn} as arising from the fact that
there can be ``special'' or ``degenerate'' regions of parameter space (potentially unstable under RG evolution) 
where some of the invariants in \eqref{CPC} vanish by themselves even though CP is not conserved.
Here we add to this understanding in the following way: We show that these special regions of parameter space correspond 
to very specific reductions in the size of the full ring of 2HDM basis invariants which have already been listed in \eqref{eq:DegCases}.
If the ring that actually needs to be discussed is known with certainty,
then we find that the number of required relations is \textit{always} in a one-to-one correspondence with the number of eliminated physical
parameters. On the other hand, if one is not strictly sure about which ring one is in (i.e.~if one cannot exclude a very specific form of parameter degeneracies), 
more general conditions, such as \eqref{CPC}, have to be stated. 
The proliferation in the number of relations is understood because they have to be sufficient also for all possible reductions of the ring. 
However, as we will show shortly, such a proliferation is not necessary if one exactly specifies which ring 
one actually is in. A completely analogous situation will arise for \Z2 and \U1 symmetries below.

Another new contribution we provide is that we can now, using relations between dependent invariants, 
also state necessary and sufficient conditions for CP1 (and, therefore, for CP conservation in general) 
solely in terms of CP-even invariants. 
This is the analogue of determining the area of the SM CKM unitarity triangle
in terms of the (all CP-even) length of its sides. 
As an aside, we are also able to state basis invariant sufficient conditions for parameter regions in which 
any single one of the above CP-odd invariants \eqref{CPC} is the decisive one, 
while the others vanish identically.
Crucial for all this is to be aware of existing relations between the invariants (syzygies) 
which hold even in the case of no global symmetry. A
general procedure of how to find and derive these syzygies was outlined in \cite[Sec.\ 6]{Trautner:2018ipq}. 
An overview of the lowest-order syzygies was provided in \cite[Tab.\ 1]{Trautner:2018ipq}.
We list all syzygies that we have used in this work in Appendix \ref{app:syzygies}.

We have already classified four ``special'' or ``degenerate'' regions in parameter space in equation~\eqref{eq:DegCases}.
In those regions, the 2HDM ring degrades to smaller rings (for detailed discussions of these smaller rings see Appendices~\ref{app:YTRing}-\ref{app:YTRingAl}).
We will go over these regions one by one now and discuss necessary and sufficient conditions for CP1 in each of them.
The degenerate region (I) is trivial, in the sense that no CP violation can take place whatsoever.

\subsubsection{Necessary and sufficient conditions for CP1 with no degeneracies}
Only if there are no parameter degeneracies, i.e.~if none of the relations in Eq.~\eqref{eq:DegCases} is realized,
then the \textit{full} 2HDM ring has to be discussed. In this case, requiring CP1 reduces the number of independent parameters by two, from
nine to eleven. The two necessary and sufficient conditions for CP1 are
\be\label{CPCnodeg}
\Jnv{1,2,1}=0=\Jnv{1,1,2}\;.
\ee
In case of no degeneracies, no other condition has to be checked.
Specifically, assuming \eqref{CPCnodeg}, one can further use the most general syzygies (without any further assumptions) to show that all other CP-odd basis invariants,
besides $\Jnv{3,3,0}$ and $\Jnv{3,0,3}$, vanish or are proportional to them.
For $\Jnv{3,3,0}$ and $\Jnv{3,0,3}$ one can further show the relations (see Appendix~\ref{app:YTRingAl} for details of the derivation)
\begin{align}\label{eq:352}
 \Jnv{3,3,0}\,\left[\Inv{0,1,1}^2-\Inv{0,2,0}\,\Inv{0,0,2}\right]~&=~0\;,\\\label{eq:325}
 \Jnv{3,0,3}\,\left[\Inv{0,1,1}^2-\Inv{0,2,0}\,\Inv{0,0,2}\right]~&=~0\;.
\end{align}
Hence, in the case of no degeneracies (in particular, excluding regions (II)-(IV)) also $\Jnv{3,3,0}=\Jnv{3,0,3}=0$ follows
and one has shown that all CP-odd invariants vanish. 

\subsubsection{Necessary and sufficient conditions for CP1 if ~\texorpdfstring{$\mathrm{Y}=0$}{Y=0}~ or ~\texorpdfstring{$\mathrm{T}=0$}{T=0}~ or ~\texorpdfstring{$\mathrm{Y}^2\mathrm{T}^2=\left(\mathrm{Y}\mathrm{T}\right)^2$}{alignment}~} 
We now discuss the degenerate regions (II)-(IV), cf.~section~\ref{sec:degeneracies}. 
As regions (II) and (III) trivially fulfill the alignment condition of region (IV), we partly treat these regions together.
Once condition (II), or (III), or (IV) is imposed, the number of independent parameters in the 2HDM ring reduces from eleven to eight, or eight, or nine, respectively, \textit{without} 
enhancing the symmetry.
We give a more detailed discussion of these sub-rings in Appendices~\ref{app:YTRing}-\ref{app:YTRingAl}.

In general, one can show that the YT-alignment condition implies
\begin{equation}
\Inv{0,1,1}^2=\Inv{0,2,0}\,\Inv{0,0,2}\qquad\Longrightarrow \qquad \Jnv{1,2,1}=0=\Jnv{1,1,2}\;.
\end{equation}
Hence, we find that in regions (II)-(IV) the condition~\eqref{CPCnodeg} is automatically fulfilled.
Consequently, again all CP-odd invariants vanish or are proportional to $\Jnv{3,3,0}$ and $\Jnv{3,0,3}$.
However, relations \eqref{eq:352} and \eqref{eq:325} are now \textit{trivially} fulfilled, hence, do not allow any conclusions on $\Jnv{3,3,0}$ or $\Jnv{3,0,3}$. 

For regions (II) and (III) where either $\mathrm{Y}=0$ or $\mathrm{T}=0$, clearly, all invariants containing them vanish, 
including in particular the CP-odd invariants. 
Hence, the sole necessary and sufficient condition for CP1 in each case is the vanishing of the respective ``opposite'' CP-odd invariant:
\begin{align}\label{eq:CPCdeg}
 &\mathrm{Region~(II):}\qquad \Jnv{3,0,3}=0\;,& &\text{or}&
 &\mathrm{Region~(III):}\qquad \Jnv{3,3,0}=0\;.&
\end{align}
This is one necessary and sufficient condition for CP1 each,
corresponding to the reduction of one parameter (from eight to seven) in agreement with Jacobi's criterion.

For region (IV), by contrast, one can use the alignment condition together with many syzygies to show the relation 
\begin{equation}
 \Jnv{3,3,0}^2\,\Inv{0,0,2}^3~=~\Jnv{3,0,3}^2\,\Inv{0,2,0}^3\;.
\end{equation}
This relation is non-trivial only in region (IV) and not for (II) or (III).
Hence, assuming no further degeneracy, one finds that in region (IV) also the invariants $\Jnv{3,3,0}$ and $\Jnv{3,0,3}$ are proportional to each other.
Without loss of generality one can, hence, pick one of them to vanish as necessary and sufficient condition for CP1. Imposing this
condition reduces the parameter by one from nine to eight.

This shows conclusively that we can for each region state a number of necessary and sufficient conditions which is one-to-one
with the number of independent parameters they eliminate. We are under the impression that this fact was known before, but never proven as clearly. 

\subsubsection{Necessary and sufficient conditions for CP1 in terms of CP-even invariants}
\label{sec:CP1primary}
This and the subsequent subsection are not relevant for the understanding of Figure~\ref{SymmetryMap} and may be skipped
by readers only interested in the reproduction of the symmetry map.

We move on to discuss necessary and sufficient conditions for CP conservation expressed solely in terms of CP-even invariants.
The corresponding relations, here, are directly obtained from the syzygies of the respective squared CP-odd invariants. 
These read 
\begin{equation}\label{eq:Syz242224}
\begin{split}
3\,\Jnv{1,2,1}^2~=&~3\,\Inv{1,1,1}^2\,\Inv{0,2,0} - 6\,\Inv{1,1,1}\,\Inv{1,2,0}\,\Inv{0,1,1} - \Inv{2,2,0}\,\Inv{0,1,1}^2 + \Inv{2,2,0}\,\Inv{0,2,0}\,\Inv{0,0,2} + 3\,\Inv{1,2,0}^2\,\Inv{0,0,2} \\ 
&+ 2\,\Inv{2,0,0}\,\Inv{0,1,1}^2\,\Inv{0,2,0} - 2\,\Inv{2,0,0}\,\Inv{0,2,0}^2\,\Inv{0,0,2}\;, \\
3\,\Jnv{1,1,2}^2~=&~3\,\Inv{1,1,1}^2\,\Inv{0,0,2} - 6\,\Inv{1,1,1}\,\Inv{1,0,2}\,\Inv{0,1,1} - \Inv{2,0,2}\,\Inv{0,1,1}^2 + \Inv{2,0,2}\,\Inv{0,0,2}\,\Inv{0,2,0} + 3\,\Inv{1,0,2}^2\,\Inv{0,2,0} \\ 
&+ 2\,\Inv{2,0,0}\,\Inv{0,1,1}^2\,\Inv{0,0,2} - 2\,\Inv{2,0,0}\,\Inv{0,0,2}^2\,\Inv{0,2,0}\;, 
\end{split}
\end{equation}
as well as
\begin{equation}
\begin{split}\label{eq:660and606}
27\,\Jnv{3,3,0}^2~=~&-\Inv{2,2,0}^3 - 54\,\Inv{1,2,0}^3\,\Inv{3,0,0} + 9\,\Inv{1,2,0}^2\,\Inv{2,0,0}^2\,\Inv{0,2,0} + 108\,\Inv{3,0,0}^2\,\Inv{0,2,0}^3 \\
   &+ 3\,\Inv{2,2,0}^2\,\Inv{2,0,0}\,\Inv{0,2,0} - 4\,\Inv{2,0,0}^3\,\Inv{0,2,0}^3 + 9\,\Inv{2,2,0}\,\Inv{1,2,0}^2\,\Inv{2,0,0} - 
   54\,\Inv{2,2,0}\,\Inv{1,2,0}\,\Inv{3,0,0}\,\Inv{0,2,0}\;, \\
27\,\Jnv{3,0,3}^2~=~&-\Inv{2,0,2}^3 - 54\,\Inv{1,0,2}^3\,\Inv{3,0,0} + 9\,\Inv{1,0,2}^2\,\Inv{2,0,0}^2\,\Inv{0,0,2} + 108\,\Inv{3,0,0}^2\,\Inv{0,0,2}^3 \\
   &+ 3\,\Inv{2,0,2}^2\,\Inv{2,0,0}\,\Inv{0,0,2} - 4\,\Inv{2,0,0}^3\,\Inv{0,0,2}^3 + 9\,\Inv{2,0,2}\,\Inv{1,0,2}^2\,\Inv{2,0,0} - 
   54\,\Inv{2,0,2}\,\Inv{1,0,2}\,\Inv{3,0,0}\,\Inv{0,0,2}\;. 
\end{split}
\end{equation}
Simply setting the left hand sides of these four equations to zero gives the necessary and sufficient 
conditions for CP conservation exclusively in terms of CP-even invariants on the right hand side. 

Note that these relations still involve secondary invariants. 
It is possible to obtain relations solely in terms of a chosen set of primary invariants by using the general syzygies, some of which are stated in \eqref{eq:Syz222}-\eqref{eq:Syz422}.
In this way all besides a chosen set of primary invariants can be eliminated.
The details of such an elimination procedure crucially depend on the choice of a set of primary invariants. 
Depending on this choice, expressions may be required to be of high order
and may become exceedingly lengthy. 
We perform this procedure in detail now for the most general case of no degeneracies above. 
The degenerate cases are much easier, as one starts in a smaller ring then, implying that many of the secondary invariants vanish or are already related.

\medskip
\textit{ Non-degenerate case.---} 
For the non-degenerate case we choose \Inv{2,0,0}, \Inv{0,2,0}, \Inv{0,0,2}, \Inv{0,1,1}, \Inv{1,2,0}, \Inv{1,0,2}, \Inv{2,1,1}, and \Inv{1,1,1} as set of algebraically independent invariants
(the procedure would already be much more complicated in case one choses \Inv{3,0,0} instead of \Inv{1,1,1}). 
As a first step we ``symmetrize'' the relations~\eqref{eq:Syz242224} by multiplying them by \Inv{0,0,2} and \Inv{0,2,0}, respectively, and sum them.
Then we can use the syzygy of the order $\mathrm{Q}^2\mathrm{Y}^2\mathrm{T}^2$, stated in \eqref{eq:Syz222}, to eliminate a combination of the invariants
$\Inv{2,2,0}$ and $\Inv{2,0,2}$ in favor of $\Inv{1,1,1}$. The resulting relation reads
\begin{equation}\label{eq:244CP}
\begin{split}
0~=~& 3\,\Inv{1,1,1}^2\,\Inv{0,2,0}\,\Inv{0,0,2} + 3\,\Inv{1,1,1}^2\,\Inv{0,1,1}^2 - 2\,\Inv{2,1,1}\,\Inv{0,1,1}^3 + 2\,\Inv{2,1,1}\,\Inv{0,1,1}\,\Inv{0,2,0}\,\Inv{0,0,2} \\
    &-6\,\Inv{1,1,1}\,\Inv{1,2,0}\,\Inv{0,1,1}\,\Inv{0,0,2} - 6\,\Inv{1,1,1}\,\Inv{1,0,2}\,\Inv{0,1,1}\,\Inv{0,2,0} + 3\,\Inv{1,2,0}^2\,\Inv{0,0,2}^2 + 3\,\Inv{1,0,2}^2\,\Inv{0,2,0}^2 \\
    &-3\,\Inv{1,2,0}\,\Inv{1,0,2}\,\Inv{0,1,1}^2 + 3\,\Inv{1,2,0}\,\Inv{1,0,2}\,\Inv{0,2,0}\,\Inv{0,0,2} \\
    &+2\,\Inv{2,0,0}\,\Inv{0,1,1}^2\,\Inv{0,2,0}\,\Inv{0,0,2}- 3\,\Inv{2,0,0}\,\Inv{0,2,0}^2\,\Inv{0,0,2}^2 + \Inv{2,0,0}\,\Inv{0,1,1}^4\;.
\end{split}
\end{equation}
As promised, it only contains our choice of primary invariants, and it will only hold in case $\Jnv{1,2,1}=\Jnv{1,1,2}=0$.
A second relation is obtained directly from the syzygy \eqref{eq:Syz233}. Again using \eqref{eq:Syz222} to eliminate $\Inv{2,2,0}$ and $\Inv{2,0,2}$ in favor of $\Inv{1,1,1}$,
and assuming the vanishing of \textit{at least} one of the invariants \Jnv{1,2,1} or \Jnv{1,1,2}, the resulting relation reads
\begin{equation}\label{eq:233CP}
\begin{split}
0~=~& 3\,\Inv{1,1,1}^2\,\Inv{0,1,1}+ \Inv{2,1,1}\,\Inv{0,1,1}^2 - \Inv{2,1,1}\,\Inv{0,2,0}\,\Inv{0,0,2} - 3\,\Inv{1,1,1}\,\Inv{1,2,0}\,\Inv{0,0,2}- 3\,\Inv{1,1,1}\,\Inv{1,0,2}\,\Inv{0,2,0} \\
  & +3\,\Inv{1,0,2}\,\Inv{1,2,0}\,\Inv{0,1,1} + 2\,\Inv{2,0,0}\,\Inv{0,2,0}\,\Inv{0,0,2}\,\Inv{0,1,1} - 2\,\Inv{2,0,0}\,\Inv{0,1,1}^3\;.
\end{split}
\end{equation}
Together, \eqref{eq:244CP} and \eqref{eq:233CP} are again two necessary and sufficient conditions for CP1 in the non-degenerate case, 
this time completely in terms of CP-even primary invariants.
After imposing CP1, there are 9 independent invariants left. A convenient choice of primary invariants to discuss the ascension from CP1 to \Z2 in the non-degenerate case, will turn out to be the 3 $\rep{1}$'s, together with
$\mathcal{I}_{2,0,0}$, $\mathcal{I}_{0,2,0}$, $\mathcal{I}_{0,0,2}$, $\mathcal{I}_{1,1,1}$, $\mathcal{I}_{0,1,1}$ and $\mathcal{I}_{2,1,1}$.

\medskip
\textit{ Special parameter regions (II) or (III).---} 
For the smaller rings, i.e.~in the special regions of parameter space, one can use an analogous procedure.
In fact, for the QY- and QT-rings (i.e.~setting either $\mathrm{T}=0$ or $\mathrm{Y}=0$) the relations in
\eqref{eq:660and606} give already the sought result: Besides the squared CP-odd invariants they only contain 
the primary invariants of the respective rings.
After imposing CP1 the number of independent invariants here is reduced by one, from six to five.
A convenient choice of non-trivial primary invariants after imposing CP1 will turn out to be $\mathcal{I}_{2,0,0}$, $\mathcal{I}_{0,2,0}$, $\mathcal{I}_{1,2,0}$ and $\mathcal{I}_{2,2,0}$
(or their respective $\mathrm{Y}\leftrightarrow\mathrm{T}$ symmetric versions) 
and we give more details on that in Appendix~\ref{app:Z23}.

\medskip
\textit{ Special parameter region (IV).---}
For the special region (IV) one may use the general syzygy for $\Jnv{3,3,0}\Jnv{3,0,3}$ stated in Eq.~\eqref{eq:Syz633}.
Making use of the many relations which arise after requiring the YT alignment condition, stated in Appendix~\ref{app:YTRingAl}, 
this syzygy simplifies to 
\begin{equation}
\begin{split}\label{eq:633}
27\,\Jnv{3,3,0}\,\Jnv{3,0,3}~=~&-\Inv{2,1,1}^3 - 54\,\Inv{1,1,1}^3\,\Inv{3,0,0} + 9\,\Inv{1,1,1}^2\,\Inv{2,0,0}^2\,\Inv{0,1,1} + 108\,\Inv{3,0,0}^2\,\Inv{0,1,1}^3 \\
   &+ 3\,\Inv{2,1,1}^2\,\Inv{2,0,0}\,\Inv{0,1,1} - 4\,\Inv{2,0,0}^3\,\Inv{0,1,1}^3 + 9\,\Inv{2,1,1}\,\Inv{1,1,1}^2\,\Inv{2,0,0} - 
   54\,\Inv{2,1,1}\,\Inv{1,1,1}\,\Inv{3,0,0}\,\Inv{0,1,1}\;.
\end{split}
\end{equation}
The vanishing of the LHS, or equivalently the RHS gives a single necessary and sufficient condition for CP1 
in region~(IV). Note the striking similarity of~\eqref{eq:633} to both equations in~\eqref{eq:660and606}, which is 
a manifestation of the similarity of the rings in regions (II)-(IV).
Again, the RHS of equation~\eqref{eq:633} already contains only primary invariants of the ring in region~(IV),
which can be chosen as \Inv{2,0,0}, \Inv{3,0,0}, \Inv{0,1,1}, \Inv{1,1,1}, and \Inv{2,1,1}.
The ring in this case contains one additional independent invariant which does not participate in the above relations
and which can be taken as either \Inv{0,2,0} or \Inv{0,0,2}, corresponding to the magnitude of Y or T, respectively.

We end this section with the following remark:
The most interesting choice for a set of primary invariants to rewrite these conditions 
certainly would be a phenomenologically motivated one, based on physical observables. However, setting the stage for
this would require to have a parameterization of the 2HDM basis invariants solely in terms of physical observables, which is not established yet. 

\subsubsection{Parameter regions of ``single-invariant dominance''}
Let us also give a basis invariant answer to the question: What are parameter regions such that exactly three out of four of the CP-odd 
invariants in \eqref{CPC} vanish? Of course, this is also an incontestable proof of the 
fact that we indeed need to check all four of those invariants to test for CPC in general.
Example points in a specific basis have been given for each of the cases, see \cite{Gunion:2005ja} and its generalization discussed in~\cite{Boto:2020wyf},
but the corresponding basis invariant relations were unknown as far as we are concerned.

It is easy to see that for $\Inv{0,2,0}=0$ (this is one-to-one with $Y_{\rep{3}}\equiv0$) only
\Jnv{3,0,3} needs to be probed because all other CP-odd invariants in \eqref{CPC} are identically vanishing.
Likewise, for the case $\Inv{0,0,2}=0$ (this is one-to-one with $Z_{\rep{3}}\equiv0$) only 
\Jnv{3,3,0} can be non-zero. 

It is harder to identify cases where all but \Jnv{1,2,1} or \Jnv{1,1,2} are zero. 
A region of parameter space where this automatically happens is 
\begin{equation}\label{eq:SID1}
 \Inv{0,1,1}~=~\Inv{1,1,1}~=~\Inv{2,1,1}~=~0\;, 
\end{equation}
together with either (for the case $\Jnv{1,1,2}\equiv0$)
\begin{equation}\label{eq:SID2}
 \Inv{2,0,2}\,\Inv{0,0,2}+3\,\Inv{1,0,2}^2-2\,\Inv{2,0,0}\,\Inv{0,0,2}^2=0\;,
\end{equation}
or (for $\Jnv{1,2,1}\equiv0$)
\begin{equation}\label{eq:SID3}
 \Inv{2,2,0}\,\Inv{0,2,0}+3\,\Inv{1,2,0}^2-2\,\Inv{2,0,0}\,\Inv{0,2,0}^2=0\;.
\end{equation}
We stress that these are certainly not the only relations amongst parameters (i.e.\ not the only regions in parameter space) that lead to the situation of single invariant dominance.
In fact, the most general relation amongst parameters to warrant single invariant dominance is simply to require the vanishing of all but one of 
the sufficient CP-odd invariants directly (which is not a ``nice'' region of parameter space, but certainly possible).
We also note that simultaneously fulfilling all three of the above conditions~\eqref{eq:SID1}-\eqref{eq:SID3} 
is sufficient for CP1, but it is not necessary.

\subsection{\texorpdfstring{$\boldsymbol{\Z2}$}{Z2} symmetry, and ascending from CP1 to \texorpdfstring{$\boldsymbol{\Z2}$}{Z2}}
\label{sec:z2}
We move on to \Z2 and discuss necessary and sufficient conditions for this symmetry.
\Z2 is a HF symmetry which can be represented by the matrix\footnote{%
There are other, physically equivalent representations for \Z2 symmetries.
They are given, for example, by taking $S$ to be one of the other two Pauli matrices $\sigma_{1,2}$.
From a basis invariant viewpoint it is clear that the resulting transformations are entirely equivalent to~\eqref{matrixS:Z2}, 
because they are related to the above matrix by basis transformations.
We have checked explicitly that any of these \Z2's, taken individually, leads to exactly the same basis invariant relations.}
\be\label{matrixS:Z2}
S = \begin{pmatrix}1 & 0 \\ 0 & -1  \end{pmatrix}\;.
\ee
Using the Jacobi criterion after imposing the \Z2 symmetry one finds that there are seven algebraically independent invariants
in agreement with the established counting of independent parameters. Thus, in principle, it should be possible to identify 
four invariant relations to eliminate four invariants from a suitably chosen set of eleven primary invariants 
to arrive at seven independent invariants.  
However, just as for the case of CP1 above, also for \Z2 symmetry parameter degeneracies can
complicate the task of matching the number of relations to the number of 
to-be eliminated parameters. We will see that it is convenient to base our conditions for \Z2 symmetry on the respective 
conditions for CP1 symmetry discussed before. Starting from CP1 is not a drawback since \Z2 will always and automatically include CP1 symmetry, see~\eqref{symtree}.
Hence, also all of the CP1 relations above are necessarily fulfilled upon requiring \Z2.
We may then state necessary and sufficient conditions for \Z2, which are a combination of the conditions for CP1 (Eq.~\eqref{CPC}), 
plus some new conditions that take us from CP1 to \Z2.
Going from CP1 to \Z2, the number of parameters is reduced by two.
The naive expectation, hence, would be that two relations on top of CP1 would be required.
However, just as for the case of CP1, where the elimination of two parameters without any assumption on possible parameter degeneracies 
required four basis invariant relations, also for \Z2 we find that there are at least three relations required if 
no assumption is made regarding the potential parameter degeneracies.

Starting from CP1, a simple set of set of necessary and sufficient conditions to obtain \Z2 symmetry without any further assumptions, is given by
\begin{align}\label{Z2:R022}
\mathcal{I}_{0,1,1}^2 ~&=~ \mathcal{I}_{0,2,0}\,\mathcal{I}_{0,0,2} \,, \\
3\,\Inv{1,2,0}^2      ~&=~ 2\,\Inv{2,0,0}\,\Inv{0,2,0}^2-\Inv{2,2,0}\,\Inv{0,2,0}\;, \label{Z2:R240} \\
3\,\Inv{1,0,2}^2      ~&=~ 2\,\Inv{2,0,0}\,\Inv{0,0,2}^2-\Inv{2,0,2}\,\Inv{0,0,2}\;. \label{Z2:R204}
\end{align}
Since we have based our conditions on the CP1 case, this also directly answers the question of how one
ascends from CP1 to \Z2.\footnote{%
Necessary and sufficient conditions for \Z2 were given by Ivanov~\cite{Ivanov:2005hg}. In the language of~\cite{Nishi:2006tg} they read:
``\Z2 symmetry holds if and only if vectors $\Vec{\text{M}}$ and $\Vec{\Lambda}$ are collinear and eigenvectors of the matrix $\tilde{\Lambda}$.''
The vectors are collinear if and only if \eqref{Z2:R022} holds, while \eqref{eq:Z2222} (or alternatively, for any of the degenerate cases, \eqref{Z2:R240} and \eqref{Z2:R204}) 
warrants that they are eigenvectors of $\tilde{\Lambda}$.}

Note that after imposing \Z2, one can identify a plethora of necessary relations that we list in Appendix~\ref{app:Z21}.
While we certainly expect that there is some combination of these \Z2 necessary relations that is also sufficient for \Z2 (without taking the step over CP1),
working this out explicitly turned out to be computationally prohibitively expensive.\footnote{
The situation could most likely be improved if one uses RGE's exclusively in terms of basis invariants.}
To confirm that the above conditions are indeed sufficient for \Z2 on top of CP1, we have checked explicitly that the conditions \eqref{Z2:R022}-\eqref{Z2:R204} and their simultaneous 
solutions are stable under RGE running using the conventional parametrization of the 2HDM scalar potential \eqref{potV} and the 1-loop RGE's stated in \cite[p.\ 153]{Branco:2011iw}.
We could not find another combination of the invariant relations stated in Appendix~\ref{app:Z21}
for which this would be true (starting from the CP1 symmetric case). 
For the interested reader we strongly recommend to read Appendix~\ref{app:Z22} now, where we 
give the detailed considerations which helped us to arrive at the above conditions.
As a remark, note that by using (i) the generally valid syzygies (which hold even in the case of no symmetry), (ii) the CP1 relations, and (iii) \eqref{Z2:R022}-\eqref{Z2:R204}, 
it \textit{must} be possible to show all of the necessary conditions for \Z2 listed in Appendix~\ref{app:Z21} by tedious but straightforward algebra 

\subsubsection{Necessary and sufficient conditions for \texorpdfstring{$\boldsymbol{\Z2}$}{Z2} with or without parameter degeneracies}
The conditions above are the most general and hold for all possible cases of parameter degeneracies. 
However, if one can for certainty say whether or not there are any of the specific parameter degeneracies in Eq.~\eqref{eq:DegCases},
then the number of necessary and sufficient conditions can be reduced, and again, as in the CP1 case, be matched
one-to-one to the number of thereby eliminated parameters.

\medskip
\textit{ Non-degenerate case.---} 
In the strictly non-degenerate case, i.e.~if none of the relations in Eq.~\eqref{eq:DegCases} holds, there are two conditions that 
are necessary and sufficient for \Z2 on top of CP1:
\begin{align}\label{eq:Z2022}
\mathcal{I}_{0,1,1}^2 ~&=~ \mathcal{I}_{0,2,0}\,\mathcal{I}_{0,0,2} \,, \\
3\,\mathcal{I}_{1,1,1}^2~&=~ 2\,\mathcal{I}_{2,0,0}\,\mathcal{I}_{0,1,1}^2-\mathcal{I}_{2,1,1}\,\mathcal{I}_{0,1,1}\,. \label{eq:Z2222}
\end{align}
These two conditions are one-to-one with exactly two eliminated parameters, and they exclusively relate primary invariants
of the CP1 case (if they are chosen as $\mathcal{I}_{2,0,0}$, $\mathcal{I}_{0,2,0}$, $\mathcal{I}_{0,0,2}$, $\mathcal{I}_{1,1,1}$, $\mathcal{I}_{0,1,1}$ and $\mathcal{I}_{2,1,1}$,
as already hinted at in section~\ref{sec:CP1primary}).

\medskip
\textit{ Special parameter region (I).---} 
We now discuss the degenerate cases starting with the most trivial case, namely degenerate parameter region (I) where $\mathrm{Q}=0$.
In this case the 2HDM ring degrades to the ring generated by the two triplets Y and T, see Appendix~\ref{app:YTRing}.
This ring has $6$ independent parameters, the three singlets plus $\Inv{0,2,0}$, $\Inv{0,0,2}$, and $\Inv{0,1,1}$.
Imposing the alignment condition on Y and T eliminates one parameter and one finds the sufficient conditions for \Z2 to be fulfilled.
However, caution is in order as, in fact, $\mathrm{Q}=0$ together with YT-alignment suffices to fulfill the conditions of a higher symmetry
to be discussed below: \U1. Hence, \Z2 symmetry is not realizable in the parameter region where $\mathrm{Q}=0$.

\medskip
\textit{ Special parameter regions (II) or (III).---} 
In cases (II) or (III) of degenerate parameter regions either Y or T vanishes. 
Starting from CP1 there is only one parameter removed if \Z2 is required. The corresponding necessary and sufficient
condition for \Z2 symmetry (on top of CP1) is equation~\eqref{Z2:R204} or \eqref{Z2:R240}, respectively.
Again one can chose primary invariants such that the relation involves only them
(in this case $\mathcal{I}_{2,0,0}$, $\mathcal{I}_{0,2,0}$, $\mathcal{I}_{1,2,0}$ and $\mathcal{I}_{2,2,0}$
or their respective $\mathrm{Y}\leftrightarrow\mathrm{T}$ conjugate versions). 
More details on the derivation of these conditions and the elimination of the other invariants
are given in Appendix~\ref{app:Z22}.

\medskip
\textit{ Special parameter regions (IV).---}
In region (IV) the YT-alignment condition is fulfilled by assumption.
Again, only one parameter is removed upon requiring \Z2 on top of CP1. 
The corresponding necessary and sufficient condition for \Z2 on top of CP1 is~\eqref{eq:Z2222}.
Again this relation can be understood as linking primary invariants exclusively, if they are chosen to contain $\mathcal{I}_{2,0,0}$, $\mathcal{I}_{0,1,1}$, $\mathcal{I}_{1,1,1}$ and $\mathcal{I}_{2,1,1}$.
The elimination of $\mathcal{I}_{3,0,0}$ works in exact analogy to cases of region (II) and (III),
described in Appendix~\ref{app:Z22}.
We remark that if there are no further parameter degeneracies, also eqs.~\eqref{Z2:R240} or \eqref{Z2:R204} would work as necessary
and sufficient conditions here, as one can then easily show their equivalence to~\eqref{eq:Z2222} 
with the considerations provided in Appendix~\ref{app:YTRingAl}. 

\medskip
Finally, note that in order to arrive at \Z2 even the strictly non-degenerate case picks up the YT-alignment condition.
Hence, the non-degenerate case merges with the degenerate parameter case (IV) at the level of \Z2 symmetry.
This is also reflected by the fact that both cases now contain the same number of independent parameters,
or primary invariants, namely seven. We now move on to see how one ascends from \Z2 to higher symmetries.

\subsubsection{From \texorpdfstring{\Z2}{Z2} to \texorpdfstring{\U1}{U(1)}}
We continue our discussion with the ascension from \Z2 to \U1,
anticipating some details of \U1 symmetry that will be discussed in detail in the subsequent section. 

Enhancing the symmetry from \Z2 to \U1 requires one additional relation. 
We continue the ascension from \Z2 to \U1 with the primary invariants that we have used 
in ascending from nothing to CP1 and from CP1 to \Z2 above.
We discuss this for the non-degenerate case, which is, upon requiring \Z2, anyways identical to 
the YT-aligned case. For the Y or T degenerate cases the discussion works completely analogous.
Starting from \Z2, we choose the non-trivial primary invariants $\mathcal{I}_{2,0,0}$, 
$\mathcal{I}_{0,1,1}$, $\mathcal{I}_{2,1,1}$, and $\mathcal{I}_{0,2,0}$ (the latter may, without loss of generality, be replaced by $\mathcal{I}_{0,0,2}$).
The necessary and sufficient condition to ascend from \Z2 to \U1 then is given by
\begin{equation}\label{Z2toU1A}
 \Inv{2,1,1}~=~-2\,\Inv{2,0,0}\,\Inv{0,1,1}\;.
\end{equation}
This can be confirmed by a straightforward algebraic computation, which shows that \eqref{Z2toU1A} together with the \Z2 conditions 
(eventually also using CP1 relations and the general syzygies) indeed implies all necessary conditions that we could
identify for the \U1 symmetry.\footnote{%
As a short convincing argument we remark that going from $\Z2$ generated by \eqref{matrixS:Z2}, to \U1 generated by \eqref{matrixS:U1} in the conventional 
parameterization requires setting $|\lambda_5|$ to zero. This is exactly what is implied by the relation \eqref{Z2toU1A} (for the already \Z2 symmetric case).}

For the $\mathrm{Y}=0$ or $\mathrm{T}=0$ degenerate cases the primary invariants at the level of \Z2 are $\mathcal{I}_{2,0,0}$, $\mathcal{I}_{0,2,0}$, and $\mathcal{I}_{2,2,0}$
(or their respective $\mathrm{Y}\leftrightarrow\mathrm{T}$ conjugated versions). Hence, the 
completely analogous necessary and sufficient conditions to ascend to \U1 from \Z2 are 
\begin{equation}\label{U1:I202b}
\mathcal{I}_{2,0,2}~=~-2\,\mathcal{I}_{2,0,0}\,\mathcal{I}_{0,0,2}\,,\qquad\text{or}\qquad\mathcal{I}_{2,2,0}~=~-2\,\mathcal{I}_{2,0,0}\,\mathcal{I}_{0,2,0}\,,
\end{equation}
respectively. Again, all \U1 relations can be shown to follow from these together with the CP1 and \Z2 relations as well as the general syzygies.

\subsubsection{From \texorpdfstring{\Z2}{Z2} to CP2: setting \texorpdfstring{\Inv{0,0,2}}{I002} and \texorpdfstring{\Inv{0,2,0}}{I020} to zero}
The invariant conditions to arrive at CP2 are rather simple: $\mathcal{I}_{0,2,0}=\mathcal{I}_{0,0,2}=0$.
In fact, CP2 can be reached by these conditions starting from any other smaller symmetry; but it cannot be reached by requiring 
any relation among existing invariants.
This suggests that instead of regarding CP2 as an ascension from \Z2 it should rather 
be regarded as a whole new starting point for a ring, namely the double degenerate case of the special parameter regions (II) and (III) together.
The five independent invariants one finds in the CP2 symmetric ring are the three trivial singlets next to $\mathcal{I}_{2,0,0}$ and $\mathcal{I}_{3,0,0}$.
Our view of CP2 as the starting point of a new ``strand'' of symmetries is supported by the fact that $\mathcal{I}_{3,0,0}$ appears here as an 
independent invariant, while in the previous ascension from 
$\text{no symmetry}\rightarrow \mathrm{CP1}\rightarrow\Z2\rightarrow\U1$ it had to be eliminated as an independent primary invariant 
already in the very first step going from no symmetry to CP1.

Nonetheless, we stress that the CP2 symmetric case can be reached from the \Z2 symmetric case by imposing another 
\Z2 symmetry on top of the existing symmetry generated by \eqref{matrixS:Z2}
\cite{Davidson:2005cw,Ferreira:2009wh}.
For example, the symmetry which is commonly called $\Pi_2$ is generated by a matrix (in the basis where \eqref{matrixS:Z2} is fixed)
\be\label{matrixS:Z2equiv}
S_2 = \begin{pmatrix} 0 & 1 \\ 1 & 0  \end{pmatrix}\;.
\ee
In the conventional parametrization this implies the parameter relations $\lambda_1=\lambda_2$, $m_{11}^2=m_{22}^2$, and $\mathrm{Im}(\lambda_5)=0$
on top of the already fulfilled \Z2 symmetry conditions $\lambda_6=\lambda_7=0$ and $m_{12}^2=0$.
This leads to a complete vanishing of the $Y_{\rep{3}}$ and $Z_{\rep{3}}$ building blocks, implying the vanishing of any basis invariant containing those.
Of course, the statement of vanishing invariants will hold in any basis. In this sense, $\Z2\times\Z2$ is not a realizable symmetry
but has a larger accidental symmetry, namely CP2. Indeed, CP2 is the smallest symmetry that enforces the exact vanishing of
the triplet building blocks. In the geometric language, CP2 is understood as a point reflection on the origin \cite{Maniatis:2007vn,Ferreira:2010yh}, 
an operation that no non-vanishing vector can be symmetric under. Hence, imposing CP2 directly leads to the exact vanishing of Y and T building blocks.

\subsection{\boldmath\U1 symmetry\unboldmath}
\label{sec:u1}
Finally we discuss the \U1 symmetry. 
One possibility to implement a \U1 HF symmetry in the 2HDM is the Peccei-Quinn symmetry, generated by
\be\label{matrixS:U1}
    S = \begin{pmatrix} e^{-i\xi} & 0 \\ 0 & e^{i\xi}  \end{pmatrix}\;,
\ee
for real values of $\xi$ (not multiples of $\pi/2$).\footnote{%
\label{foot:U1}%
There are other, physically equivalent possibilities for \U1 symmetries in the 2HDM
which are generated, for example, by the exponentiation of either of the other two Pauli matrices $\sigma_{1,2}$. 
From a basis invariant viewpoint it is clear that the resulting one-parameter transformations are entirely equivalent to \eqref{matrixS:U1}, 
because they are related to the above matrix by basis transformations. We have checked explicitly that any of these \U1's, taken individually, 
leads to exactly the same basis invariant relations.}
Imposing this symmetry, the Jacobi criterion informs us that there should be six independent invariants.
Of course, this number six corresponds to the number of six physical parameters of the 2HDM scalar sector with \U1 symmetry shown in \cite{Branco:2011iw}.
Consequently, in the non-degenerate case one would expect that there should be five relations among the set of $11$ independent invariants stated above.
Given our previous discussion, a straightforward way to state these conditions is to combine all of the above conditions (2 for CP1, 2 for \Z2 and 1 for \U1).
For the non-degenerate case these five relations would be given by $\Jnv{3,3,0}=\Jnv{3,0,3}=0$ together with Eqs.~\eqref{eq:Z2022}, \eqref{eq:Z2222} and \eqref{Z2toU1A} 
(note that alignment implies that these are actually only four independent relations, which is sufficient since alignment itself eliminates two parameters and so 
we can go from 11 to 6 parameters with only four relations). 
However, this is not the most elegant way to state the necessary and sufficient conditions for \U1 directly. 
Using the general syzygies one can show the equivalence of these conditions to the relations
\begin{align}\label{eq:U1600}
 \Inv{3,0,0}^2~=&~\left(\tfrac{1}{3}\,\Inv{2,0,0}\right)^3\;,\\ \label{eq:U1022}
 \Inv{0,1,1}^2~=&~~\Inv{0,2,0}\,\Inv{0,0,2}\;,\\ \label{eq:U1211}
 \Inv{2,1,1}~=&~-2\,\Inv{2,0,0}\,\Inv{0,1,1}\;,\\ \label{eq:U311}
 \Inv{2,0,0}\,\Inv{1,1,1}~=&~-6\,\Inv{3,0,0}\,\Inv{0,1,1}\;.
\end{align}
These are the complete necessary and sufficient conditions for \U1 in the non-degenerate case.

For the degenerate case with $\mathrm{Q}=0$ already the YT-alignment condition itself is necessary and sufficient for \U1.
For the degenerate cases with $\mathrm{Y}=0$ or $\mathrm{T}=0$ one needs three conditions (which eliminate three parameters), 
namely \eqref{eq:U1600} together with 
\begin{align}
 \Inv{2,2,0}~=&~-2\,\Inv{2,0,0}\,\Inv{0,2,0}\;,\\
 \Inv{2,0,0}\,\Inv{1,2,0}~=&~-6\,\Inv{3,0,0}\,\Inv{0,2,0}\;,
\end{align}
or their respective $\mathrm{Y}\leftrightarrow\mathrm{T}$ conjugate versions.

Finally, we also state necessary and sufficient conditions for \U1 that work for \textit{all} parameter regions irrespective of any parameter degeneracies.
We find that the minimal number of such relations is six, and they can be stated as\footnote{%
A necessary and sufficient criterion for a global \U1 symmetry has been given in~\cite{Ivanov:2005hg}.
Stated in the language of~\cite{Nishi:2006tg} it reads ``the PQ symmetry holds, if and only if two eigenvalues of $\tilde{\Lambda}$ coincide, while
the vectors $\Vec{\text{M}}$ and $\Vec{\Lambda}$ are eigenvectors of $\tilde{\Lambda}$ corresponding to the non-degenerate third eigenvalue''.
As already discussed in footnote~\ref{foot:CP3}, \eqref{U1:I300} corresponds to two degenerate eigenvalues in $\tilde{\Lambda}$.
The alignment of vectors $\Vec{\text{M}}$ and $\Vec{\Lambda}$ corresponds to \eqref{U1:I011}. 
The final part of the criterion, namely that the vectors correspond to the non-degenerate eigenvalue of $\tilde{\Lambda}$, 
is in a basis invariant way expressed as relations \eqref{U1:I220}, \eqref{U1:I202}, \eqref{U1:I320} and \eqref{U1:I302}.}
\begin{align}\label{U1:I300}
 \Inv{3,0,0}^2~=&~\left(\tfrac{1}{3}\,\Inv{2,0,0}\right)^3\;,\\\label{U1:I011}
 \Inv{0,1,1}^2~=&~~\Inv{0,2,0}\,\Inv{0,0,2}\;,\\ \label{U1:I220}
 \Inv{2,2,0}~=&~-2\,\Inv{2,0,0}\,\Inv{0,2,0}\;,\\ \label{U1:I202}
 \Inv{2,0,2}~=&~-2\,\Inv{2,0,0}\,\Inv{0,0,2}\;,\\ \label{U1:I320}
 \Inv{2,0,0}\,\Inv{1,2,0}~=&~-6\,\Inv{3,0,0}\,\Inv{0,2,0}\;, \\ \label{U1:I302}
 \Inv{2,0,0}\,\Inv{1,0,2}~=&~-6\,\Inv{3,0,0}\,\Inv{0,0,2}\;.
\end{align}
For the (non-)degenerate cases these reduce to the (four)three relations above, warranting that the 
number of relations in any specific case is always one-to-one with the number of eliminated parameters,
just as for CP1 and \Z2 above.
The set of six algebraically independent invariants at the level of \U1, therefore, can be chosen as 
$Y_{\rep{1}}$, $Z_{\rep{1}_{(1)}}$, $Z_{\rep{1}_{(2)}}$,  $\mathcal{I}_{2,0,0}$,  $\mathcal{I}_{0,2,0}$, and $\mathcal{I}_{0,0,2}$.

\subsubsection{From \texorpdfstring{\U1}{U(1)} to CP3: setting \texorpdfstring{\Inv{0,0,2}}{I 002} and \texorpdfstring{\Inv{0,2,0}}{I 020} to zero}
Again it is instructive to see how one can ascend from the \U1 symmetry to more symmetric cases.
We note that the relation between $\mathcal{I}_{2,0,0}$ and  $\mathcal{I}_{3,0,0}$ stated in \eqref{U1:I300} 
is precisely the one for the CP3 model \eqref{CP3:relation},
related to the basis invariant $D$ of Ref.~\cite{Ferreira:2009wh}.
The difference between \U1 and CP3, thus, lays exclusively 
in the (non-)vanishing of the triplet building blocks: If both of the triplet building blocks, and therefore also the invariants \Inv{0,0,2} or \Inv{0,2,0},
are identically zero one ascends from \U1 to CP3. 
The simultaneous vanishing of both of these invariants can be enforced by the transformation $\Phi_1\leftrightarrow\Phi_2$ (in the basis relative to \eqref{matrixS:U1}), 
commonly denoted as $\Pi_2$. Imposing $\Pi_2$ on top of the \U1 leaves us with exactly the same 
four algebraically independent invariants as in the CP3 case
\cite{Ferreira:2009wh}.  We stress that if only one of the invariants $\mathcal{I}_{0,2,0}$ or $\mathcal{I}_{0,0,2}$ is zero, we are still only in the \U1 symmetric 
case. Nonetheless, a special situation arises that we will briefly discuss now.

\subsubsection{Distinction of seemingly interchangeable invariants at the quantum level}
\label{sec:InvsandRGE}
Note that our whole discussion of the construction of invariants and their symmetry-based relations is
entirely symmetric under the exchange of $Z_{\rep{3}}\leftrightarrow Y_{\rep{3}}$ (i.e.~exchange of T and Y). 
This ``symmetry'' originates from the fact that these building blocks transform identically under basis changes.
However, this apparent ``symmetry'' is broken by quantum effects.

\renewcommand{\arraystretch}{1.5}
\begin{table}[t]
\begin{center}
\begin{tabular}{cccccc}
\toprule[1pt]
\centering
\multirow{2}{*}{Symmetry} & \multirow{2}{*}{~~$n_p$~~} & \multirow{3}{*}{\shortstack{Choice of algebraically \\ independent invariants \\ (not unique)}} & \multicolumn{3}{c}{Sym.\ enhancement upon requiring:} \\
& & & \multirow{2}{*}{$Y^2\rightarrow0 \land T^2\rightarrow0$}~~ & ~~\multirow{2}{*}{$Q^3\sim\,Q^2$ (\text{Eq.}\ref{CP3:relation})}~~ & ~~\multirow{2}{*}{Eqs.~\eqref{Z2:R022}-\eqref{Z2:R204}} \\ \\
\hline
\SU2 & 3 &  - & -  & -  & -  \\
CP3 & 4 &   $Q^2$ & -  &-  & - \\
CP2 & 5 &  $Q^2$, $Q^3$ & - & CP3 & -  \\
\U1 & 6 &  $Q^2$, $Y^2$, $T^2$ & CP3 & - & -  \\
$\mathbb{Z}_2$ & 7 &  $Q^2$, $Y^2$, $T^2$, $Q^2YT$ & CP2 & U(1) & - \\
CP1 & 9 &  $Q^2$, $Y^2$, $T^2$, $YT$, $QYT$, $Q^2YT$ & CP2 & - & $\mathbb{Z}_2$   \\
no symmetry & 11 &  $Q^2$, $Y^2$, $T^2$, $YT$, $QY^2$, $QT^2$, $QYT$, $Q^2YT$ & CP2 & - & -   \\
\bottomrule[1pt]
\end{tabular}
\end{center}
\caption{\label{tab:invariants}
The first two columns show the global symmetry of the Higgs potential and corresponding number of independent parameters $n_p$. 
In the subsequent column we show a choice of (algebraically) independent basis invariants for each case, 
which for all cases also includes the three trivial singlet invariants \eqref{singlets} which are not displayed. 
In the last three columns we show how the symmetries are enhanced (likewise, the number of independent parameters is reduced) 
if specific relations are required for, or among, the respective invariants. 
}
\end{table}
Let us explain this in more detail. If only one of the invariants $\Inv{0,0,2}$ or $\Inv{0,2,0}$ is zero, the global symmetry of the Higgs potential is not increased. 
Hence, one would expect that a zero value of only one of the triplet building blocks should be 
a scale dependent statement, i.e.\ a non-zero value for the vanishing invariant should be re-generated by quantum corrections.
Let us express the invariants explicitly to explore this in more detail.\footnote{%
Expressing the invariants in an explicit parametrization would not be necessary if we had used the renormalization group equations expressed 
directly in terms of the invariants, see~\cite{Bednyakov:2018cmx}.}
We use App.\ B and D of \cite{Trautner:2018ipq} and employ the conventional
parametrization of the 2HDM scalar potential \eqref{potV}.
Upon enforcing the \U1 symmetry the invariants read\footnote{%
In fact, the \Z2 symmetry discussed in section~\ref{sec:z2} is already 
sufficient to enforce this form of the invariants.}
\begin{align}\label{I002:U1}
\mathcal{I}_{0,0,2}\, &\stackrel{\U1}{=}  \, \frac{1}{16}(\lambda_2-\lambda_1)^2\;,\\\label{I020:U1}
\mathcal{I}_{0,2,0}\, &\stackrel{\U1}{=}  \, \frac{1}{4}(m_{11}^2-m_{22}^2)^2\;.
\end{align}
If and only if these invariants vanish, also the corresponding building blocks vanish. 
That is, if in addition to the presence of the \U1, also $\lambda_1=\lambda_2$ and/or $m_{11}^2=m_{22}^2$ 
then all components of $Z_{\rep{3}}$ and/or $Y_{\rep{3}}$ are zero. 

Now setting only one of the invariants \eqref{I002:U1} or \eqref{I020:U1} to zero does \textit{not} lead to an enhanced symmetry,
implying that a single zero invariant should be regenerated by renormalization group running.
The one-loop renormalization equations for the relevant couplings are given by\footnote{%
We ignore gauge and Yukawa couplings here which do, in general, also contribute. 
}\cite[p.\ 153]{Branco:2011iw}
\begin{align}
 16\pi^2\frac{\mathrm{d}}{\mathrm{d}\ln\mu}\left(\lambda_1-\lambda_2\right)~=~&
 12\left(\lambda_1^2-\lambda_2^2\right)\;,\\
 16\pi^2\frac{\mathrm{d}}{\mathrm{d}\ln\mu}\left(m_{11}^2-m_{22}^2\right)~=~& 
 -\left(m_{11}^2-m_{22}^2\right)\left(4\,\lambda_3+2\,\lambda_4\right)+6\,\lambda_1\,m_{11}^2-6\,\lambda_2\,m_{22}^2\;.
\end{align}
We observe that \Inv{0,2,0}, if zero at some scale, is regenerated from the quartic couplings by 
renormalization group equation (RGE) effects, as expected
(unless we are at the symmetric point $\lambda_1=\lambda_2$).
However, a zero invariant \Inv{0,0,2} is \textit{not} regenerated.
Of course, this is the well-known hallmark of the fact that quadratic operators can ``softly'' break CP3,
while a breaking of CP3 by quartic terms is ``hard''. 
The novel insight here is the following: Despite the fact that the construction of invariants,
as well as the symmetry relations of invariants above and below, 
are all completely symmetric under the exchange of $Z_{\rep{3}}\leftrightarrow Y_{\rep{3}}$,
this exchange ``symmetry'' is broken by quantum effects.
That is, despite the fact that basis transformations act absolutely identically on $Z_{\rep{3}}$ and $Y_{\rep{3}}$,
renormalization effects do distinguish them. This distinction then is also propagated to all invariants constructed from them.
Despite the fact that all unphysical basis changes (which can, in a basis-dependent treatment also obscure renormalization, see e.g.\ \cite{Bednyakov:2018cmx}) 
do cancel in the invariants, renormalization effects will be differing depending on the mass dimension (and anomalous dimension) of each invariant.
This quantum-level distinction of the invariants is an issue worth further investigation, beyond the scope of this work.

\subsection{Summary and discussion}
\label{sec:summary}
\renewcommand{\arraystretch}{1.5}
\begin{table}[t]
\begin{center}
\begin{tabular}{ccc}
\toprule[1pt]
\centering
Symmetry & \multirow{2}{*}{\shortstack{Necessary and sufficient conditions \\ on basis invariants}} & \\ \\
\hline
CP1 & $\Jnv{1,2,1}=0\;,\;\;\Jnv{1,1,2}=0\;,\;\;\Jnv{3,3,0}=0\;,\;\;\Jnv{3,0,3}=0\;,$ \\
\hline
\multirow{4}{*}{\Z2} & $\Jnv{3,3,0}=0\;,\;\;\Jnv{3,0,3}=0\;,$ \\
                     & $\mathcal{I}_{0,1,1}^2 = \mathcal{I}_{0,2,0}\,\mathcal{I}_{0,0,2}\;,$ \\
                     & $3\,\Inv{1,2,0}^2     = 2\,\Inv{2,0,0}\,\Inv{0,2,0}^2-\Inv{2,2,0}\,\Inv{0,2,0}\;,$\qquad $3\,\Inv{1,0,2}^2    = 2\,\Inv{2,0,0}\,\Inv{0,0,2}^2-\Inv{2,0,2}\,\Inv{0,0,2}\;,$ \\
\hline
\multirow{4}{*}{\U1} & $\Inv{0,1,1}^2=\Inv{0,2,0}\,\Inv{0,0,2}\;,$ \\ 
                     & $\Inv{3,0,0}^2=\left(\tfrac{1}{3}\,\Inv{2,0,0}\right)^3\;,$ \\
                     & $\Inv{2,2,0}=-2\,\Inv{2,0,0}\,\Inv{0,2,0}\;,$ \qquad  $\Inv{2,0,2}=-2\,\Inv{2,0,0}\,\Inv{0,0,2}\;,$ \\
                     & $\Inv{2,0,0}\,\Inv{1,2,0}=-6\,\Inv{3,0,0}\,\Inv{0,2,0}\;,$ \qquad $\Inv{2,0,0}\,\Inv{1,0,2}=-6\,\Inv{3,0,0}\,\Inv{0,0,2}\;,$ \\
\midrule[1pt]
CP2 & $\Inv{0,2,0}=0\;,$ \qquad $\Inv{0,0,2}=0\;,$ \\
\hline
\multirow{2}{*}{CP3} & $\Inv{0,2,0}=0\;,$ \qquad $\Inv{0,0,2}=0\;,$ \\
                     & $\Inv{3,0,0}^2=\left(\tfrac{1}{3}\,\Inv{2,0,0}\right)^3\;,$ \\
\midrule[1pt]
\multirow{2}{*}{\SU2}& $\Inv{0,2,0}=0\;,$ \qquad $\Inv{0,0,2}=0\;,$ \\
                     & $\Inv{2,0,0}=0\;,$ \qquad $\Inv{3,0,0}=0\;.$ \\
\bottomrule[1pt]
\end{tabular}
\end{center}
\caption{\label{tab:summary}
Necessary and sufficient conditions for each of the six classes of global symmetries of the most general 2HDM scalar potential.
The conditions are ``fail-proof'' in the sense that no other conditions have to be checked whatsoever, i.e.\ the conditions apply to \textit{all} cases, 
also if parameters of the potential are potentially degenerate. Of course, in order to check whether or not a given symmetry is realized, one still has to check the conditions of the next higher symmetry,
as smaller symmetries are implied by the higher symmetries according to Eq.~\eqref{symtree}.}
\end{table}
The main summary of how to ascend and descend amongst the possible symmetries of the most general 2HDM has already been 
shown in Figure~\ref{SymmetryMap}. In addition, in Table~\ref{tab:invariants} we give the same information in a tabular form.
For convenience, in Table~\ref{tab:summary} we summarize the necessary and sufficient conditions 
for each class of symmetries in a ``fail-proof'' way, i.e.\ such that no extra conditions on parameter degeneracies etc.\ need to be checked
whatsoever.
The conditions are necessary and sufficient for each given symmetry. However, one should still check
the conditions of the next higher symmetry in the ordering of Eq.~\eqref{symtree}, to make sure that 
not actually a higher symmetry is realized.
In this form, the conditions could easily be implemented, for example, in a computer code to automatize the detection of symmetries
irrespectively of the chosen basis. 
While for experimental predictions this form is perhaps of limited use, our approach is very useful for the theoretical detection 
of symmetries (and approximate symmetries) from measurements, as well as for the conceptual understanding of how global symmetries are related to the 
algebraic structure of a potential. 

In this sense, one of our main results here is not specific to the 2HDM, 
but regards our new, conceptually unprecedented method of analyzing physical systems in terms of the ring of systematically constructed basis invariant quantities.
While one of our original expectations was that one might ``see'' what symmetries are possible directly from the basis invariants,
this hope was not fulfilled. Nonetheless, we learned a great deal, namely the very important distinction between symmetries
that can be reached  by the interrelation (or factorization) of basis invariants (i.e.~leaving a ring ``intact''),
and symmetries that can only be reached if certain building blocks of invariants are forced to be absent (i.e.~``collapsing'' the ring to a smaller ring). 
This distinction, which has to the best of our knowledge never been made, is of uttermost importance, not only for this specific model but in general.
Furthermore, we have learned that it is generally hard to decide whether a given relation between invariants
is sufficient to warrant an enhanced symmetry; many of the relations one finds by imposing symmetries 
are only necessary. The ultimate check here is the (all order) RGE stability of a given set of invariant relations,
which makes sense, as symmetries are inherently non-perturbative quantum mechanical statements.

While in this work we have used the route from symmetries to their basis invariant relations,
it would also be very desirable to revert this route, i.e.~to deduce symmetries starting from 
basis invariant relations. 
This direction requires more research, and one has to admit that with our approach presented here 
it is currently impossible to tell which symmetries are possible.
In this sense we benefited a lot from the vast knowledge in the literature about the possible, realizable symmetries of the 2HDM.
We expect that this situation could be improved a lot, if one would know the RGEs directly in terms of invariants.
Eventually, this would allow one to deduce sensible relations between invariants based on their behavior on the system
of coupled RGEs. In this respect it seems noteworthy that our relations among basis invariants \textit{always} related 
building blocks with themselves. That is, the ``building block number'' is conserved in all the equations,
meaning that relations like $Y=T$ or $Q^2=T^2$ \textit{etc.} never appear.

In view of this, the most promising route to generally determine possible realizable symmetries still seems to be 
to determine all possible alignments of basis-covariant objects \cite{Ivanov:2019kyh}.
This is then related to questions of the type: ``to what subgroups can one break \SU2 with two triplets and a five-plet'',
which have been studied to some extend (see e.g.\ \cite{Adulpravitchai:2009kd,Merle:2011vy,Luhn:2011ip}) but are known to be a hard problem in general.
The algebraic approach to this question is both, not well developed and not very well known in the field.

\section{Conclusions}\enlargethispage{1.5cm}
\label{sec:conclusions}

We have derived necessary and sufficient conditions for all realizable global symmetries of the most general 2HDM in terms of relations between 
basis invariants.  These conditions are collected in Table~\ref{tab:summary}, formulated in terms of the basis invariants defined in \eqref{eq:invariants} 
(we state some of the invariants in the conventional parametrization in Appendix~\ref{app:InvariantsInConvPar}, and refer to~\cite[Appendix D]{Trautner:2018ipq}
where all of them are stated in a parametrization independent notation).

Furthermore, we have clarified how one can ascend or descend between the different classes of symmetries and this is summarized in the ``Symmetry Map'' of the model, Figure~\ref{SymmetryMap}.
In deriving this, we have found that there are, in general, two algebraically very different ways how symmetries are realized: 
On the one hand, basis invariant objects can be non-trivially related, in which case the number of parameters is reduced one-to-one with respect to the number of 
relations imposed. On the other hand, basis covariant building blocks of the invariants can vanish (i.e.\ can be forced to vanish by symmetry), 
which leads to the vanishing of all invariants containing them. This changes the structure of the ring of basis invariants and can reduce the number of 
parameters by more than the number of relations required.
Regions in the parameter space that have previously been called ``special'' or ``degenerate'' 
are identified as exactly those regions where certain basis covariant objects vanish, or are aligned in such a way that essentially corresponds to
a vanishing.

If no assumption is made about the exact structure of the ring of invariants (i.e.\ if one does not want to preclude that 
one or the other building block might eventually vanish) then the number of conditions that are necessary and sufficient for a given symmetry
is typically greater than the number of eliminated parameters. For the 2HDM this was known to be the case for CP1 symmetry,
and here we have shown that it is also true for \Z2 and \U1 symmetries. Complementarily, we have also shown that if one is absolutely sure
about which covariants vanish or not, then one can always find necessary and sufficient conditions for a symmetry whose number is one-to-one with
the number of eliminated parameters. 

We note that our results have largely been derived by using the known realizable symmetries of the 2HDM to 
deduce necessary invariant relations, and subsequently using the renormalization group running to identify sufficient relations.
The other way around, i.e.\ starting from the invariants to deduce symmetries and the respective relations that lead to them is 
a very different open problem that we did not address here. 
Nontheless, we expect the present study certainly will be a useful start to this problem as well. 

Other important results presented here for the first time are an in-depth discussion of the rings and sub-rings contained in 
the 2HDM parameter space, as well as several of the 2HDM syzygies (relations between invariants that hold irrespective of any symmetry).
Furthermore we have shown necessary and sufficient conditions for CP conservation in the 2HDM solely in terms of CP-even invariants.

On more general grounds, we have seen that on a purely algebraic level there is an exchange ``symmetry'' among identically transforming basis covariant 
building blocks and their constructed invariants (here $Z_{\rep{3}}\leftrightarrow Y_{\rep{3}}$ or equivalently $\mathrm{T}\leftrightarrow\mathrm{Y}$).
Nontheless, the covariants and the invariants constructed from them can be distinguished by quantum effects.

Finally, we stress that the methodology and ideas used in this paper are completely general and apply to other models as well,
with the future most obvious application being the SM flavor sector.

\section*{Acknowledgments}
The work of M.P.B., R.B. and J.P.S. was supported by the Portuguese
Funda\c{c}\~{a}o para a Ci\^{e}ncia e a Tecnologia (FCT) through the
projects UID/FIS/00777/2019, PTDC/FIS-PAR/29436/2017, and
CERN/FIS-PAR/0004/2017; these projects are partially funded through POCTI
(FEDER), COMPETE, QREN, and the EU. M.P.B. is also supported by FCT under
the project SFRH/BD/146718/2019. 
The work of A.T.\ has partly been supported by a postdoc 
fellowship of the German Academic Exchange Service (DAAD).

\appendix
\section{Invariants in conventional parametrization}
\label{app:InvariantsInConvPar}
For convenience we explicitly state here some of the invariants given in Appendix B and D of \cite{Trautner:2018ipq} in the conventional parameterization of the 2HDM scalar potential, Eq.~\eqref{potV}, 
in a basis where\footnote{%
Such a basis can always be chosen following \cite{Gunion:2005ja}. We use this basis here to obtain short explicit expressions for the invariants.
We stress that, of course, none of our basis invariant statements depends on any particular choice of basis.}
$\lambda_7=-\lambda_6$.
\begin{align}
 Y_{\text{\textbf{1}}}~=&~ m_{11}^2+m_{22}^2\;, \\
 Z_{\text{\textbf{1}}_{(1)}}~=&~\frac{1}{2} (\lambda_1+\lambda_2+\lambda_3+\lambda_4)\;, \\
 Z_{\text{\textbf{1}}_{(2)}}~=&~\frac{\lambda_3-\lambda_4}{2}\;, \\
 \mathcal{I}_{0,2,0}~=&~\Re(m_{12}^2)^2+\Im(m_{12}^2)^2+\frac{1}{4} (m_{11}^2-m_{22}^2)^2\;, \\
 \mathcal{I}_{0,0,2}~=&~\frac{1}{4} (\lambda_1-\lambda_2)^2\;, \\
 \mathcal{I}_{0,1,1}~=&~\frac{1}{4} (\lambda_1-\lambda_2) (m_{11}^2-m_{22}^2)\;, \\
 \mathcal{I}_{2,0,0}~=&~\frac{1}{12} \left[\lambda_1+\lambda_2-2 (\lambda_3+\lambda_4)\right]^2
                       +\Re(\lambda_5)^2+\Im(\lambda_5)^2+4\left[\Re(\lambda_6)^2+\Im(\lambda_6)^2\right]\;, \\ 
 \mathcal{I}_{1,2,0}~=&~-\frac{1}{6} \left[\lambda_1+\lambda_2-2 (\lambda_3+\lambda_4)\right] \left[\Re(m_{12}^2)^2+\Im(m_{12}^2)^2-\frac12 (m_{11}^2-m_{22}^2)^2\right]+\\\notag
  &\Re(\lambda_5) \Re(m_{12}^2)^2-\Re(\lambda_5)\Im(m_{12}^2)^2+2\,\Im(\lambda_5)\Re(m_{12}^2)\Im(m_{12}^2)+\\\notag
  &2\left(-m_{11}^2+m_{22}^2\right)\left[\Im(\lambda_6)\Im(m_{12}^2)+\Re(\lambda_6)\Re(m_{12}^2)\right]\\
 \mathcal{I}_{1,0,2}~=&~\frac{1}{12} (\lambda_1-\lambda_2)^2 \left[\lambda_1+\lambda_2-2 (\lambda_3+\lambda_4)\right]\;\\
 \mathcal{I}_{3,0,0}~=&~-\frac{1}{216}\left[\lambda_1+\lambda_2-2 (\lambda_3+\lambda_4)\right]^3+
 2\,\Re(\lambda_5) \Re(\lambda_6)^2-2\,\Re(\lambda_5)\Im(\lambda_6)^2-4\,\Im(\lambda_5)\Im(\lambda_6)\Re(\lambda_6)+ \\\notag
 &\frac{1}{6} \left[\lambda_1+\lambda_2-2 (\lambda_3+\lambda_4)\right] \left[\Re(\lambda_5)^2+\Im(\lambda_5)^2-2\left(\Re(\lambda_6)^2+\Im(\lambda_6)^2\right)\right]\;, \\
 \mathcal{I}_{1,1,1}~=&~\frac12 (-\lambda_1+\lambda_2) \left[2\,\Im\left(\lambda_6\right) \Im\left( m_{12}^2\right)+2 \Re\left(\lambda_6\right) \Re\left( m_{12}^2\right)
                                 -\frac16 (m_{11}^2-m_{22}^2)\left[\lambda_1+\lambda_2-2 (\lambda_3+\lambda_4)\right]\right]\;,\\
 \mathcal{I}_{2,1,1}~=&~\frac{1}{2} (-\lambda_1+\lambda_2)\left\{ -\left[\lambda_1+\lambda_2-2\left(\lambda_3+\lambda_4\right)\right] 
                        \left[\Re(\lambda_6) \Re(m_{12}^2)+\Im(\lambda_6) \Im(m_{12}^2)-\frac{1}{12}(m_{11}^2-m_{22}^2)\right]\right.+\\\notag
 &6 \,\Re(\lambda_5)\Im( \lambda_6)\Im(m_{12}^2)-6 \,\Im(\lambda_5)\Re( \lambda_6)\Im(m_{12}^2)-6 \,\Im(\lambda_5)\Im( \lambda_6)\Re(m_{12}^2)-6 \,\Re(\lambda_5)\Re( \lambda_6)\Re(m_{12}^2)+\\\notag
 &\left.(m_{11}^2-m_{22}^2) \left[2\left(\Re(\lambda_6)^2-\Im(\lambda_6)^2\right)-\Re(\lambda_5)^2-\Im(\lambda_5)^2\right]\right\}\;.
\end{align}

\section{The 2HDM ring and its reductions}
\subsection{The full 2HDM ring}
\label{app:QYTring}
The full 2HDM ring is constructed from the building blocks $\mathrm{Q}$, $\mathrm{Y}$, and $\mathrm{T}$
and their covariant transformation properties under changes of basis (see \cite{Trautner:2018ipq} for more details).
The multi-graded Hilbert series is given by
\begin{equation}
HS(q,y,t)=\frac{N\left(q,y,t\right)}{D\left(q,y,t\right)}, 
\end{equation}
with numerator
\begin{equation}
\begin{split}
N\left(q,y,t\right)~=~&1+q t y+q^2 t y+q t^2 y+q t y^2+q^2 t^2 y+q^2 t y^2+q^3 t^3+q^3 t^2 y+q^3 t y^2+q^3 y^3-q^3 t^4 y-q^3 t^3 y^2 \\
&-q^3 t^2 y^3-q^3 t y^4-q^4 t^3 y^2-q^4 t^2 y^3-q^5 t^3 y^2-q^5 t^2 y^3-q^4 t^3 y^3-q^5 t^3 y^3-q^6 t^4 y^4\;,
\end{split}
\end{equation}
and denominator
\begin{equation}
D\left(q,y,t\right)~=~\left(1-t^2\right) \left(1-y^2\right) \left(1-t y\right) \left(1-q^2\right) \left(1-q^3\right) \left(1-q t^2\right) \left(1-q y^2\right)\left(1-q^2t^2\right) \left(1-q^2y^2\right)\;.
\end{equation}
This leads to the ungraded Hilbert series
\begin{equation}\label{eq:ungradedHS}
HS(z)~\equiv~HS(q=z,y=z,t=z)~=~\frac{1+z^3+4\,z^4+2\,z^5+4\,z^6+z^7+z^{10}}{\left(1-z^2\right)^4\left(1-z^3\right)^3\left(1-z^4\right)}\;.
\end{equation}

\subsection{The 2HDM ring if \texorpdfstring{$\boldsymbol{\mathrm{Q}=0}$}{Q=0}}
\label{app:YTRing}
We briefly discuss the structure of the 2HDM ring in the special region of parameter space where $\mathrm{Q}=0$, that is the 
$\rep{5}$-plet building block identically vanishes. We note that 
\begin{equation}
 \Inv{2,0,0}=0\,,\quad \Inv{3,0,0}=0\, \qquad \Longrightarrow \qquad \mathrm{Q}=0\;.
\end{equation}
In the opposite direction, the vanishing of Q implies that all invariants that contain Q are identically zero.
Hence, the only remaining building blocks are Y and T, and the so-called YT-ring is constructed for two $\rep{3}$-plets of \SU2.
The multi-graded Hilbert series of the YT-ring is given by
\be
HS(y,t)=\frac{1}{\left(1-y^2\right)\left(1-t^2\right)\left(1-y t\right)},
\ee
and the plethystic logarithm is
\be
PL(q,y)=y^2+t^2+y t.
\ee
This ring is free.
The complete set of invariants is 
\be
\mathcal{I}_{0,2,0},\quad \mathcal{I}_{0,0,2},\quad \mathcal{I}_{0,1,1}\;.
\ee
Geometrically, the basis invariant information of two triplet vectors of \SU2 corresponds to their
lengths, as well as their relative angle; this is exactly what these invariants correspond to.

\subsection{The 2HDM ring if \texorpdfstring{$\boldsymbol{\mathrm{Y}=0}$}{Y=0} \textit{or} \texorpdfstring{$\boldsymbol{\mathrm{T}=0}$}{T=0}}
\label{app:QYRing}
Here we consider separately the ``QY'' and ``QT'' sectors,
which are obtained from the complete ring of 2HDM invariants by setting $\mathrm{T}=0$ or $\mathrm{Y}=0$,
respectively. It makes sense to consider the QY and QT rings together in one go, because both, T and Y, 
correspond to objects that transform identically (namely, as triplets) under the \SU2 basis change. 
For definiteness we treat the QY-ring here.
In all expressions one may substitute Y for T and they stay equally valid.
We note that
\begin{equation}
 \Inv{0,2,0}=0\quad\Longrightarrow\quad\mathrm{Y}=0\;, \qquad{\text{and}}\qquad \Inv{0,0,2}=0\quad\Longrightarrow\quad\mathrm{T}=0\;,
\end{equation}
while $\mathrm{T}=0$ or $\mathrm{Y}=0$ imply the vanishing of all invariants containing them.
Hence, setting $\mathrm{T}=0$, the full set of invariants in \eqref{setIs} and \eqref{generatingSet} 
is reduced to  six invariants, namely 
\be\label{eq:QYring}
\mathcal{I}_{2,0,0},\quad \mathcal{I}_{0,2,0},\quad \mathcal{I}_{3,0,0},\quad \mathcal{I}_{1,2,0},\quad \mathcal{I}_{2,2,0},\quad \mathcal{J}_{3,3,0}\;.
\ee
The multi-graded Hilbert series of the QY-ring (ring of a $\rep{5}$-plet and a single $\rep{3}$-plet of \SU2) is given by
\be
HS(q,y)=\frac{1+q^3y^3}{\left(1-q^2\right)\left(1-y^2\right)\left(1-q^3\right)\left(1-qy^2\right)\left(1-q^2y^2\right)},
\ee
and the plethystic logarithm is
\be
PL(q,y)=q^2+y^2+q^3+qy^2+q^2y^2+q^3y^3-q^6y^6.
\ee
This implies that out of the six invariants~\eqref{eq:QYring} only five are independent, and there is a relation
of the structure $\mathrm{Q}^6\mathrm{Y}^6$. This relation has already been stated in \eqref{eq:660and606},
which shows the dependence of the (squared) CP-odd invariant $\mathcal{J}_{3,3,0}$ on the other five invariants.

The separate considerations of QY and QT rings will be useful and very instructive in order to understand 
how we arrived at the sufficient conditions for the \Z2 symmetry stated in Sec.~\ref{sec:z2}.
We give more details on that in Appendix~\ref{app:Z22}.

\subsection{The 2HDM ring if \texorpdfstring{$\boldsymbol{\mathrm{Y}\sim\mathrm{T}}$}{YsimT}}
\label{app:YTRingAl}
In the situation where Y and T are aligned (i.e. condition (IV) in Eq.~\eqref{eq:DegCases} holds) the 2HDM ring is, in principle, also reduced to a ring of 
a single $\rep{5}$-plet and a single $\rep{3}$-plet of \SU2. Hence, the discussion of the previous section applies. 
However, compared to the situation there, here we find one additional independent invariant that simply corresponds to 
the length of the second vector. However, there is no new ``geometric'' information or new possibility of symmetry breaking 
involved in this ring as compared to the QY or QT rings.
We note that the YT-alignment condition $\Inv{0,1,1}^2=\Inv{0,2,0}\Inv{0,0,2}$ (together with the general syzygies) implies the following relations 
\begin{align}
 \Inv{1,2,0}\,\Inv{0,0,2}~&=~\Inv{1,0,2}\,\Inv{0,2,0}\;, \\
 \Inv{2,2,0}\,\Inv{0,0,2}~&=~\Inv{2,0,2}\,\Inv{0,2,0}\;, \\
 \Inv{1,1,1}^2~&=~\Inv{1,2,0}\,\Inv{1,0,2}\;, \\
 \Inv{2,1,1}^2~&=~\Inv{2,2,0}\,\Inv{2,0,2}\;.
\end{align}
We may use those to eliminate invariants from the generating set (i.e.~show their dependence).
A set of independent primary invariants in this case may be chosen as 
\be
\mathcal{I}_{2,0,0},\quad \mathcal{I}_{3,0,0},\quad \mathcal{I}_{0,2,0},\quad \mathcal{I}_{0,0,2},\quad \mathcal{I}_{1,1,1},\quad \mathcal{I}_{2,1,1}\;,
\ee
and for the (dependent) CP-odd invariant one may chose either $\Jnv{3,3,0}$ or $\Jnv{3,0,3}$, while noting that they are related via (this expression holds only in the aligned case)
\begin{equation}
 \Jnv{3,3,0}^2\,\Inv{0,0,2}^3~=~\Jnv{3,0,3}^2\,\Inv{0,2,0}^3\;.
\end{equation}
For completeness we also note the following important fact: The alignment condition is \textit{sufficient} for the vanishing of \Jnv{1,2,1} and \Jnv{1,1,2}.
Importantly, without any further assumption on alignment or not, $\Jnv{1,2,1}=\Jnv{1,1,2}=0$ together with the general syzygies
(syzygies that we have used to arrive at the relations below are collected in Eq.~\eqref{eq:Syz442}-\eqref{eq:Syz332II})
can be used to show that also $\Jnv{2,2,1}=0$, and $\Jnv{2,1,2}=0$, and ultimately the relations 
\begin{align}
 \Jnv{3,2,1}\,\Inv{0,1,1}~&=~\Jnv{3,1,2}\,\Inv{0,2,0}\;,\\
 \Jnv{3,1,2}\,\Inv{0,1,1}~&=~\Jnv{3,2,1}\,\Inv{0,0,2}\;,\\[0.2cm]
 3\,\Jnv{3,3,0}\,\Inv{0,1,1}~&=~\Jnv{3,2,1}\,\Inv{0,2,0}\;,\\
 3\,\Jnv{3,0,3}\,\Inv{0,1,1}~&=~\Jnv{3,1,2}\,\Inv{0,0,2}\;,\\[0.2cm]
 3\,\Jnv{3,3,0}\,\Inv{0,0,2}~&=~\Jnv{3,2,1}\,\Inv{0,1,1}\;,\\
 3\,\Jnv{3,0,3}\,\Inv{0,2,0}~&=~\Jnv{3,1,2}\,\Inv{0,1,1}\;.
\end{align}
Together these imply 
\begin{align}
 \Jnv{3,3,0}\,\left[\Inv{0,1,1}^2-\Inv{0,2,0}\,\Inv{0,0,2}\right]~&=~0\;,\\
 \Jnv{3,0,3}\,\left[\Inv{0,1,1}^2-\Inv{0,2,0}\,\Inv{0,0,2}\right]~&=~0\;.
\end{align}
This proves the known fact that if the alignment condition holds, then CP may be violated, 
but only via the invariants \Jnv{3,3,0} and \Jnv{3,0,3}.
However, note that our investigation also implies the other direction: 
If CP is violated but $\Jnv{1,2,1}=\Jnv{1,1,2}=0$ holds, then the alignment condition \textit{must} be fulfilled.

\subsection{The 2HDM ring if \texorpdfstring{$\boldsymbol{\mathrm{Y}=0}$}{Y=0} \textit{and} \texorpdfstring{$\boldsymbol{\mathrm{T}=0}$}{T=0}}
\label{app:QRing}
In case both $\mathrm{T}=0$ and $\mathrm{Y}=0$ the only building block is the $\rep{5}$-plet Q.
The multi-graded Hilbert series of the Q-ring is given by
\be
HS(q,y)=\frac{1}{\left(1-q^2\right)\left(1-q^3\right)},
\ee
and the plethystic logarithm is
\be
PL(q,y)=q^2+q^3.
\ee
This ring is free. The full set of invariants is 
\be\label{eq:Qring}
\mathcal{I}_{2,0,0}\quad\text{and}\quad\mathcal{I}_{3,0,0}\;.
\ee
Intuitively, it is clear that a $\rep{5}$-plet of \SU2, corresponding to a real traceless symmetric matrix, has two basis invariant degrees of freedom
corresponding to two independent eigenvalues.

\section{Necessary relations for the \texorpdfstring{\Z2}{Z2} symmetric case}
\label{app:Z21}
Here we show basis invariant relations which are fulfilled when the potential is \Z2 symmetric, 
i.e.~all these relations are necessary for \Z2.
\begin{align}
\mathcal{I}_{0,1,1}^2 ~&=~ \mathcal{I}_{0,2,0}\,\mathcal{I}_{0,0,2}\,, \label{Z2:I022}\\ 
\mathcal{I}_{1,1,1}^2 ~&=~ \mathcal{I}_{1,0,2}\,\mathcal{I}_{1,2,0}\,, \label{Z2:I222I} \\
\mathcal{I}_{1,2,0}\,\mathcal{I}_{0,0,2} ~&=~ \mathcal{I}_{1,0,2}\,\mathcal{I}_{0,2,0}\,,\label{Z2:I122}\\
\mathcal{I}_{2,2,0}\,\mathcal{I}_{0,0,2} ~&=~ \mathcal{I}_{2,0,2}\,\mathcal{I}_{0,2,0}\,,\label{Z2:I222II}\\
\mathcal{I}_{2,1,1}\,\mathcal{I}_{0,1,1} ~&=~ 2\,\mathcal{I}_{2,0,0}\,\mathcal{I}_{0,1,1}^2-3\,\mathcal{I}_{1,0,2}\,\mathcal{I}_{1,2,0}\,, \label{Z2:I222III}\\ 
2\,\Inv{3,0,0}\,\mathcal{I}_{0,1,1}^3 ~&=~ \mathcal{I}_{1,1,1}\left(\mathcal{I}_{0,1,1}^2\mathcal{I}_{2,0,0}-\mathcal{I}_{1,0,2}\mathcal{I}_{1,2,0}\right)\,, \label{Z2:666}
\end{align}
\begin{align}\label{eq:QYZ2conditions}
 0~=&~3\,\Inv{1,2,0}^2-2\,\Inv{2,0,0}\,\Inv{0,2,0}^2+\Inv{2,2,0}\,\Inv{0,2,0}\;,& &\left[240\right]& &\cancel{\Inv{3,0,0}}& \\\label{eq:QYZ2conditionsSECOND}
 0~=&~2\,\Inv{3,0,0}\,\Inv{0,2,0}^3+\Inv{1,2,0}^3-\Inv{2,0,0}\,\Inv{1,2,0}\,\Inv{0,2,0}^2\;,& &\left[360\right]\mathrm{I}& &\cancel{\Inv{2,2,0}}& \\
 0~=&~4\,\Inv{3,0,0}\,\Inv{0,2,0}^3-\Inv{1,2,0}^3-\Inv{2,2,0}\,\Inv{1,2,0}\,\Inv{0,2,0}\;,& &\left[360\right]\mathrm{II} & &\cancel{\Inv{2,0,0}}& \\
 0~=&~108\,\Inv{3,0,0}^2\,\Inv{0,2,0}^3+\Inv{2,2,0}^3-3\Inv{2,0,0}^2\,\Inv{2,2,0}\,\Inv{0,2,0}^2-2\,\Inv{2,0,0}^3\,\Inv{0,2,0}^3\;,& &\left[660\right]& &\cancel{\Inv{1,2,0}}& \\\label{eq:QYZ2conditionsLAST}
 0~=&~54\,\Inv{1,2,0}^3\,\Inv{3,0,0}^2-9\,\Inv{1,2,0}^2\,\Inv{2,2,0}\,\Inv{2,0,0}\,\Inv{3,0,0}-\Inv{1,2,0}^3\,\Inv{2,0,0}^3& &\left[960\right]& &\cancel{\Inv{0,2,0}}& \\\notag
    &~+ \,\Inv{1,2,0}\,\Inv{2,2,0}^2\,\Inv{2,0,0}- \Inv{2,2,0}^3\,\Inv{3,0,0}\;.& && &&
\end{align}
Note that the last set of five relations contains exclusively Q and Y, not T. 
Analogous relations hold for the formal replacement of $\mathrm{Y}\leftrightarrow\mathrm{T}$ (corresponding to $\Inv{q,i,j}\leftrightarrow\Inv{q,j,i}$ in the other notation) 
but we do not display them.
In the second and third column of the last set of relations we also state the structure of the relations in powers of Q,Y,T, and the respective invariant 
that does not appear in the relation. This will be important for the discussion in Appendix~\ref{app:Z22}.
We strongly suspect that some combination of these relations should also be sufficient for \Z2 symmetry (starting from no symmetry and/or starting from CP1).
Only four of all these relations are expected to be really algebraically independent.

\section{Derivation of sufficient relations for the \texorpdfstring{\Z2}{Z2} symmetric case}
\label{app:Z22}
Here we provide some extra considerations which are very instructive 
in order to understand how we arrived at the sufficient conditions for the \Z2 symmetry stated in Sec.~\ref{sec:z2}.
For this, consider separately the ``QY'' and ``QT'' sectors,
which are obtained from the complete ring of 2HDM invariants by setting $\mathrm{T}=0$ or $\mathrm{Y}=0$,
respectively. It makes sense to consider these sectors separately because both, T and Y, 
correspond to objects that transform identically (namely, as triplets) under the \SU2 basis change. 
Hence, the possibilities of 
having a symmetric alignment of Q and Y, or Q and T are the same.
In other words, having \SU2 broken to subgroups by a frozen five-plet and a frozen triplet representation,
or by a frozen five-plet and two frozen triplet representations, the symmetry may never be enhanced by adding the
second frozen triplet. On the other hand, one may, of course, reduce the symmetry 
by adding a second triplet, \textit{if} it is not aligned with the first one. 
Hence the symmetry situation in the full ring will be the largest common symmetry of the QY and QT-rings, 
plus an eventual symmetry reduction stemming from a relative misalignment of Y and T that can be 
grasped by a bunch of ``link'' relations which care about the relative alignment of Y and T.

Without loss of generality we treat the QY-ring here. In all expressions one may substitute Y for T and they stay equally valid.
Details of the invariants and the Hilbert series of this ring have already been discussed in Appendix~\ref{app:QYRing}.
We now move on to discuss necessary and sufficient conditions for enhanced symmetries in the QY-ring. 

Quite obviously, the necessary and sufficient condition for CP conservation in this ring is $\mathcal{J}_{3,3,0}=0$.
Requiring this, implies an additional relation between the remaining five invariants, namely \eqref{eq:660and606} with $\mathrm{LHS}=0$. 
Hence, the number of algebraically independent invariants is reduced to four.
However, which four algebraically independent invariants we choose is, as always, completely up to us.

We move on to the next possible higher symmetry, namely \Z2.
Upon imposing \Z2 symmetry the number of independent invariants must further be reduced from four to three;
implying the advent of a new, independent relation among our choice of four remaining invariants.
Since the choice of four remaining invariants after imposition of CP is not fixed, we may try to simply
leave out one-by-one of the invariants in \eqref{eq:QYring} and see whether we can identify a relation among
the remaining invariants. Indeed, after imposing \Z2 symmetry, we can find five relations where each one of the relations is independent of 
one of the invariants. These relations have already been stated in Eqs.~\eqref{eq:QYZ2conditions}-\eqref{eq:QYZ2conditionsLAST} above.
All of these relations are necessary for conservation of a \Z2 symmetry in the QY-ring. 
However, it turns out that not all of these conditions are sufficient. There are multiple ways to check whether or not a given condition is sufficient:
\begin{enumerate}
 \item[(1.)] A very rudimentary (and often prohibitively difficult) way to check, at least, whether a given invariant relation is \textit{excluded} from being sufficient is: 
 (i) Pick an explicit parameterization for the invariants; (ii) Solve the condition for one of the parameters; 
 (iii) Plug the corresponding solution back into all other \Z2-necessary conditions and check if they are fulfilled as well.  
 \item[(2.)] Check whether a given invariant relation is sufficient to entirely eliminate one of the remaining four 
 independent primary invariants.
 \item[(3.)] Check the RGE stability of a given invariant relation (or respectively, of its solution). If a condition is sufficient for an enhanced symmetry
 it (or respectively, its solution) must be stable under RGE running to all orders.
\end{enumerate}

To check whether any of the conditions \eqref{eq:QYZ2conditions}-\eqref{eq:QYZ2conditionsLAST} \textit{could} be sufficient for \Z2 symmetry
we have started by using the first and most rudimentary approach (1.).
To make this feasible, we have used the conventional parametrization of Eq.~\eqref{potV} and made use of the fact that one can always chose a 
basis in which $\lambda_6=-\lambda_7$ \cite{Gunion:2005ja}. Furthermore, we have required all parameters to be real (as CP1, and hence CP is implied by \Z2),
which also amounts to a special basis choice. 
Then we find that \textit{all besides} the relation $\left[240\right]$ (Eq.~\eqref{eq:QYZ2conditions})
allow for solutions which do \textit{not} automatically fulfill the other conditions. 
Hence, this immediately excludes all but the relation $\left[240\right]$ from being sufficient conditions for a \Z2 symmetry, even though
they certainly are necessary. In other words: from the rudimentary consideration (1.) alone we are sure that for each relations besides $\left[240\right]$,
there is a region of parameters that allows to fulfill the respective relation, but which does \textit{not} require an enhanced symmetry.
While this excludes the other relations, this by itself does not warrant that $\left[240\right]$ is indeed sufficient for \Z2 symmetry.
This is the case, because we have solved $\left[240\right]$ only in a very special region of parameter space, and other, non-sufficient solution might 
exist in other regions. Nonetheless, we can use the two other ways outlined above to convince ourselves that $\left[240\right]$, together with CP symmetry, 
is indeed sufficient for \Z2 in the QY-ring. 

Following method (2.) above, we need to check whether we can use $\left[240\right]$ and CP invariance to entirely eliminate a primary invariant from our set of 
still four independent invariants. If this succeeds, it appears to us that this is, in fact, a very general proof of the sufficiency, even though we 
cannot prove this intuition mathematically. 
A possible way of elimination is as follows: There are two crucial observations involved: 1) we can solve $\left[240\right]$ for $\Inv{1,2,0}^2$, and 
$\left[240\right]$ does not contain $\Inv{3,0,0}$. 2) If we reorganize \eqref{eq:660and606} (with $\mathcal{J}_{3,3,0}=0$ imposed due to CP symmetry)
such that all terms with $\Inv{3,0,0}$ appear on one side, then the other side contains $\Inv{1,2,0}$ only in the form $\Inv{1,2,0}^2$:
\begin{align}\label{eq:replacerelation}
  & 54\,\Inv{3,0,0}\left( 2\,\Inv{3,0,0}\,\Inv{0,2,0}^3 - \Inv{1,2,0}^3 - \Inv{2,2,0}\,\Inv{1,2,0}\,\Inv{0,2,0}\right)~=~ &\\ \notag
  &\Inv{2,2,0}^3 - 9\,\underline{\Inv{1,2,0}^2}\,\Inv{2,0,0}^2\,\Inv{0,2,0} - 3\,\Inv{2,2,0}^2\,\Inv{2,0,0}\,\Inv{0,2,0} + 4\,\Inv{2,0,0}^3\,\Inv{0,2,0}^3 - 9\,\Inv{2,2,0}\,\underline{\Inv{1,2,0}^2}\,\Inv{2,0,0} \;.
\end{align}
Now we recall that we are free in our choice of independent primary invariants. That is, instead of $\Inv{3,0,0}$, we could simply use $\Inv{3,0,0}^2$
or even, and this is part of the trick, the whole LHS of \eqref{eq:replacerelation}. A quick check of the Jacobi criterion confirms that 
taking as a primary invariant the LHS of above equation instead of $\Inv{3,0,0}$, there are still five independent invariants 
in the case of no symmetry.
Hence, upon imposing CP, which implies we set $\mathcal{J}_{3,3,0}=0$, we gain \eqref{eq:replacerelation}, which can now simply be used 
on the set of independent invariants in order to eliminate one of them (namely, exactly the LHS of \eqref{eq:replacerelation} -- and $\Inv{3,0,0}$ is completely eliminated).
One arrives at four independent invariants: $\mathcal{I}_{2,0,0}$, $\mathcal{I}_{0,2,0}$, $\mathcal{I}_{1,2,0}$ and $\mathcal{I}_{2,2,0}$.

To ascend further to the \Z2 symmetric case requires imposing an additional relation.
We take relation $\left[240\right]$ (Eq.~\eqref{eq:QYZ2conditions}) in the form
\begin{equation}
3\,\Inv{1,2,0}^2~=~2\,\Inv{2,0,0}\,\Inv{0,2,0}^2-\Inv{2,2,0}\,\Inv{0,2,0}\;.
\end{equation}
Then starting with $\Inv{1,2,0}^2$ instead of $\Inv{1,2,0}$ as primary invariant,
we can seamlessly eliminate $\Inv{1,2,0}$ from the list of invariants. 
This leaves us with three independent invariants: $\mathcal{I}_{2,0,0}$, $\mathcal{I}_{0,2,0}$ and $\mathcal{I}_{2,2,0}$.
We stress that a similar elimination procedure is not possible with any of the other relations in \eqref{eq:QYZ2conditionsSECOND}-\eqref{eq:QYZ2conditionsLAST};
which might of course be due to the reason that they are not sufficient to warrant \Z2 symmetry.
It is not clear to us according to which logic one could immediately spot the ``most basic'' (i.e.\ a sufficient) relation here.
We observe in the present case, that it is the smallest relation (counting total powers of Q and Y) that does the job here, but 
it is not clear whether this is a generic feature.

Ultimately we have also used method (3.) above, to confirm that indeed, relation [240] (Eq.~\eqref{eq:QYZ2conditions}) is the only one that is 
stable under RGE running, which is the only criterion we really trust at the moment.

\medskip 

Finally, we extend the considerations from the separate QY and QT-rings to the whole 2HDM QYT-ring.
To identify sufficient conditions for \Z2 in the full ring, we adopt the conditions $\left[240\right]$ and $\left[204\right]$ which are sufficient for \Z2 in the QY and QT-ring, respectively, 
plus one (a priori potentially several) ``link'' condition that govern the relative alignment of the two triplets Y and T.
We find that it is again the smallest possible link relation that works, namely Eq.~\eqref{Z2:R022}. 
Together, [240], [204] and Eq.~\eqref{Z2:R022} (on top of CP1) provide sufficient conditions for \Z2 symmetry in the full ring.

We stress that the complications which require four conditions for the CP1 case (even though we only eliminate two parameters), and three additional conditions for the \Z2 case (again eliminating only two parameters)
do not arise in the QY and QT-rings separately. For the separate QY and QT rings, where there is only a single triplet building block, one condition eliminates one parameter in each step from no symmetry to CP1, and from CP1 to \Z2.
That is, the whole complication in the full QYT-ring arises from the possibility of having an additional relative (mis-)alignment of the triplets Y and T, or in other words, 
from having multiple copies of the same representation as building blocks.

\medskip 

For completeness, we note that we have also sought for necessary and sufficient conditions for \Z2 (still in the QY-ring) directly, without imposing CP1 to begin with.
Empirically, we have found that the conditions [240] and [360]I, or alternatively [240] and [360]II are actually sufficient for CP conservation and \Z2 symmetry in the QY-ring.
This fuels the hope that one could also state a small number of relations which are sufficient for \Z2 in the whole 2HDM ring, 
without requiring CP1 first. However, we could not show sufficiency for any set of conditions because the corresponding (highly non-linear) equations become hard 
to solve in an explicit parametrization. Also method (2.) above becomes tedious as expressions become exceedingly lengthy.
In this respect we note that we could not even proof in this way that the conditions we state in the main text (Eq.~\eqref{Z2:R204}-\eqref{Z2:R022}) 
are sufficient for \Z2. The way the proof there was conducted is via methods (1.) and (3.) above, for which things 
turn out to be much easier once CP is conserved. 
We expect all this to become easier once RGE's directly in terms of basis invariants become available, 
which is beyond the scope of this work, but should become available in the future.

\section{A solvable set of primary invariants in the \texorpdfstring{\Z2}{Z2} symmetric case}
\label{app:Z23}
For completeness, we show in this Appendix a set of primary invariants in the \Z2 symmetric case 
which allows us to express all other invariants in terms of them.
Let us chose as primary invariants the three trivial invariants together with 
$\mathcal{I}_{1,0,2}$, $\mathcal{I}_{1,2,0}$, $\mathcal{I}_{2,0,0}$ and $\mathcal{I}_{0,1,1}$.
With this set of invariants in mind, one can show the relations
\ba\label{Z2:I020}
\mathcal{I}_{0,2,0}^2\,\mathcal{I}_{1,0,2} &=& \mathcal{I}_{0,1,1}^2\,\mathcal{I}_{1,2,0}\,,\\\label{Z2:I002}
\mathcal{I}_{0,0,2}^2\,\mathcal{I}_{1,2,0} &=& \mathcal{I}_{0,1,1}^2\,\mathcal{I}_{1,0,2}\,,\\\label{Z2:I211}
\mathcal{I}_{2,1,1}\,\mathcal{I}_{0,1,1}   &=& 2\,\mathcal{I}_{0,1,1}^2\,\mathcal{I}_{2,0,0} - 3\,\mathcal{I}_{1,0,2}\,\mathcal{I}_{1,2,0}\,,\\\label{Z2:I300}
4\,\mathcal{I}_{3,0,0}^2\,\mathcal{I}_{0,1,1}^6 &=& \mathcal{I}_{1,0,2}\,\mathcal{I}_{1,2,0}\left(\mathcal{I}_{0,1,1}^2\mathcal{I}_{2,0,0}-\mathcal{I}_{1,0,2}\mathcal{I}_{1,2,0}\right)^2\,.
\ea
This allows for the elimination of the four invariants on the LHS that are not included in our choice of primaries. 
We have also derived the additional relations among the invariants of the generating set \eqref{generatingSet}, which shows that 
we can express all of them in terms of our primary invariants. While all CP-odd $\mathcal{J}$-invariants vanish, 
for the remaining three invariants one can show
\ba\label{Z2:I111}
\mathcal{I}_{1,1,1}^2  & = & \mathcal{I}_{1,0,2}\,\mathcal{I}_{1,2,0}\,,\\\label{Z2:I202}
\mathcal{I}_{2,0,2}^2\,\left(\mathcal{I}_{1,2,0}\,\mathcal{I}_{0,1,1}^2\right)  & = & \mathcal{I}_{1,0,2}\left(3\,\mathcal{I}_{1,0,2}\,\mathcal{I}_{1,2,0}\,-\,2\,\mathcal{I}_{0,1,1}^2\,\mathcal{I}_{2,0,0}\right)^2\,,\\\label{Z2:I220}
\mathcal{I}_{2,2,0}^2\,\left(\mathcal{I}_{1,0,2}\,\mathcal{I}_{0,1,1}^2\right)  & = & \mathcal{I}_{1,2,0}\left(3\,\mathcal{I}_{1,2,0}\,\mathcal{I}_{1,0,2}\,-\,2\,\mathcal{I}_{0,1,1}^2\,\mathcal{I}_{2,0,0}\right)^2\,.
\ea
We observe that in contrast to the invariant relations of symmetries higher than \Z2, which always include continuous symmetries in the chain $\U1\subset\mathrm{CP}3\subset\SU2$,
the invariant relations here contain sums of invariants, while the continuous-symmetry relations above were always purely multiplicative.
Whether or not this is a general feature (perhaps related to the continuous vs.\ discrete nature of the implied symmetries) or only by coincidence is not clear 
to us and remains to be investigated.

We stress once more that primary invariants could also be chosen as the three trivial invariants together with $\mathcal{I}_{0,0,2}$, $\mathcal{I}_{0,2,0}$, $\mathcal{I}_{2,0,0}$, and $\mathcal{I}_{3,0,0}$,
which appears to be more convenient for the ascension to higher symmetries. However, for this choice of invariants we could not come
up with expressions that solve all other invariants in terms of them. 

\section{Syzygies}
\label{app:syzygies}
Here we collect syzygies for the 2HDM invariant ring. These have been derived according to the general procedure
outlined in \cite[Sec.\ 6]{Trautner:2018ipq}. An overview of the lowest-order syzygies is provided in \cite[Tab.\ 1]{Trautner:2018ipq}.
The lowest-order syzygy is of the order $\mathrm{Q}^2\mathrm{Y^2}\mathrm{T^2}$ and it is given by
\begin{equation}\label{eq:Syz222}
 3\,\Inv{1,1,1}^2~=~2\,\Inv{2,1,1}\,\Inv{0,1,1} - \Inv{2,2,0}\,\Inv{0,0,2} - \Inv{2,0,2}\,\Inv{0,2,0} + 3\,\Inv{1,2,0}\,\Inv{1,0,2}+\Inv{2,0,0}\,\Inv{0,2,0}\,\Inv{0,0,2}-\Inv{2,0,0}\,\Inv{0,1,1}^2\;.
\end{equation}
Then there are two syzygies of the order seven, a CP-even and a CP-odd one. The CP-odd one is not of interest here, but we note that it has already been stated in \cite[Eq.\ (55)]{Trautner:2018ipq}.
The CP-even syzygy of order seven is of the structure $\mathrm{Q}^3\mathrm{Y^2}\mathrm{T^2}$ and it reads
\begin{equation}\label{eq:Syz322}
\begin{split}
 2\,\Inv{2,1,1}\,\Inv{1,1,1}~=&~\Inv{2,0,2}\,\Inv{1,2,0} + \Inv{2,2,0}\,\Inv{1,0,2} - 6\,\Inv{3,0,0}\,\Inv{0,1,1}^2 + 6\,\Inv{3,0,0}\,\Inv{0,2,0}\,\Inv{0,0,2} \\
 &- 2\,\Inv{1,0,2}\,\Inv{2,0,0}\,\Inv{0,2,0} - 2\,\Inv{1,2,0}\,\Inv{2,0,0}\,\Inv{0,0,2} + 4\,\Inv{1,1,1}\,\Inv{2,0,0}\,\Inv{0,1,1}\;.
\end{split}
\end{equation}
Furthermore, there is a relation of order eight with the structure $\mathrm{Q}^4\mathrm{Y^2}\mathrm{T^2}$ which reads
\begin{equation}\label{eq:Syz422}
\begin{split}
 \Inv{2,1,1}^2~=&~\Inv{2,0,2}\,\Inv{2,0,0}\,\Inv{0,2,0} + \Inv{2,2,0}\,\Inv{2,0,0}\,\Inv{0,0,2} +\Inv{2,0,2}\,\Inv{2,2,0} - 2\,\Inv{2,1,1}\,\Inv{0,1,1}\,\Inv{2,0,0} \\
 &- 18\,\Inv{3,0,0}\,\Inv{1,0,2}\,\Inv{0,2,0}- 18\,\Inv{3,0,0}\,\Inv{1,2,0}\,\Inv{0,0,2}+ 36\,\Inv{3,0,0}\,\Inv{1,1,1}\,\Inv{0,1,1} - \Inv{2,0,0}^2\,\Inv{0,1,1}^2 + \Inv{2,0,0}^2\,\Inv{0,2,0}\,\Inv{0,0,2}\;.
\end{split}
\end{equation}
All three of the above relations are valid in general for the 2HDM without any symmetry assumption.
One should think of these relations as \textit{each} being the reason for which one invariant is removed from the 
set of algebraically independent invariants. Note that none of these relations involve CP-odd invariants.

Next we look for two relations that involve also CP-odd basis invariants, but which do not entirely vanish upon requiring CP conservation (i.e.\ relations which involve CP-odd invariants, but only in even powers).
The first of such relations is of order eight and has the structure $\mathrm{Q}^2\mathrm{Y^3}\mathrm{T^3}$. It reads
\begin{equation}\label{eq:Syz233}
\begin{split}
3\,\Jnv{1,2,1}\,\Jnv{1,1,2}~=&~9\,\Inv{1,1,1}^2\,\Inv{0,1,1} + 2\,\Inv{2,0,2}\,\Inv{0,1,1}\,\Inv{0,2,0} + 2\,\Inv{2,2,0}\,\Inv{0,1,1}\,\Inv{0,0,2} - 3\,\Inv{2,1,1}\,\Inv{0,1,1}^2 \\
&- \Inv{2,1,1}\,\Inv{0,2,0}\,\Inv{0,0,2} - 3\,\Inv{1,1,1}\,\Inv{1,2,0}\,\Inv{0,0,2} - 3\,\Inv{1,1,1}\,\Inv{1,0,2}\,\Inv{0,2,0} - 3\,\Inv{1,2,0}\,\Inv{1,0,2}\,\Inv{0,1,1}\;.
\end{split}
\end{equation}
Finally, we look at the relation of the squared lowest-order CP-odd invariants. 
The first of such relations is of order eight and has the structure $\mathrm{Q}^2\mathrm{Y^4}\mathrm{T^2}$. It is given by
\begin{equation}\label{eq:Syz242}
\begin{split}
3\,\Jnv{1,2,1}^2~=&~3\,\Inv{1,1,1}^2\,\Inv{0,2,0} - 6\,\Inv{1,1,1}\,\Inv{1,2,0}\,\Inv{0,1,1} - \Inv{2,2,0}\,\Inv{0,1,1}^2 + \Inv{2,2,0}\,\Inv{0,2,0}\,\Inv{0,0,2} + 3\,\Inv{1,2,0}^2\,\Inv{0,0,2} \\ 
&+ 2\,\Inv{2,0,0}\,\Inv{0,1,1}^2\,\Inv{0,2,0} - 2\,\Inv{2,0,0}\,\Inv{0,2,0}^2\,\Inv{0,0,2}\;. 
\end{split}
\end{equation}
As this is not $\mathrm{Y}\leftrightarrow\mathrm{T}$ symmetric, we also take the 
corresponding relation of the structure $\mathrm{Q}^2\mathrm{Y^2}\mathrm{T^4}$ and add it to \eqref{eq:Syz242} after multiplying them by a suitable factor $\mathrm{T^2}$ or $\mathrm{Y^2}$, respectively.
The resulting relation is of the structure $\mathrm{Q}^2\mathrm{Y^4}\mathrm{T^4}$ and given by
\begin{equation}\label{eq:Syz244}
\begin{split}
3\,\Jnv{1,2,1}^2\,\Inv{0,0,2} + 3\,\Jnv{1,1,2}^2\,\Inv{0,2,0} ~=&~ 6\,\Inv{1,1,1}^2\,\Inv{0,2,0}\,\Inv{0,0,2} - 6\,\Inv{1,1,1}\,\Inv{1,2,0}\,\Inv{0,1,1}\,\Inv{0,0,2} - 6\,\Inv{1,1,1}\,\Inv{1,0,2}\,\Inv{0,1,1}\,\Inv{0,2,0} \\
&+ \Inv{2,2,0}\,\Inv{0,2,0}\,\Inv{0,0,2}^2 + \Inv{2,0,2}\,\Inv{0,2,0}^2\,\Inv{0,0,2} - \Inv{2,2,0}\,\Inv{0,1,1}^2\,\Inv{0,0,2}- \Inv{2,0,2}\,\Inv{0,1,1}^2\,\Inv{0,2,0} \\
&+ 3\,\Inv{1,2,0}^2\,\Inv{0,0,2}^2 + 3\,\Inv{1,0,2}^2\,\Inv{0,2,0}^2 + 4\,\Inv{2,0,0}\,\Inv{0,1,1}^2\,\Inv{0,2,0}\,\Inv{0,0,2} - 4\,\Inv{2,0,0}\,\Inv{0,2,0}^2\,\Inv{0,0,2}^2\;.
\end{split}
\end{equation}
These relations are generally valid, and they show the algebraic dependence of the lowest CP-odd invariants.
Note that in the case of CP conservation, the CP-odd invariants on the left-hand sides in equations \eqref{eq:Syz233} and \eqref{eq:Syz244} vanish.
This implies that there are then two new, independent relations between the CP-even basis invariants. This will reduce the number
of algebraically independent invariants by two (one for each new independent relation).

Since these relations are the simplest syzygies derivable, we suppose that also the corresponding relations between the remaining non-vanishing invariants
are the simplest relations that can be obtained. 
Note that the relations still involve invariants which may not be part of a chosen set of primary invariants. 
In order to obtain the relations solely in terms of primary invariants we can use the general syzygies, Eq.\ \eqref{eq:Syz222}-\eqref{eq:Syz422}, to eliminate dependent invariants.
Choosing the set of algebraically independent invariants as \Inv{2,0,0}, \Inv{0,2,0}, \Inv{0,0,2}, \Inv{0,1,1}, \Inv{1,2,0}, \Inv{1,0,2}, \Inv{2,1,1}, and \Inv{1,1,1} this replacement 
is actually straightforward (it is more complicated in the case that one choses \Inv{3,0,0} instead of \Inv{1,1,1}).
The resulting relations amongst the primaries, which are fulfilled only if CP is conserved, have already been stated in \eqref{eq:244CP} and \eqref{eq:233CP}. 

\medskip

Finally we also list the syzygies that allowed us to arrive at the results of Appendix~\ref{app:YTRingAl}.
\begin{equation}\label{eq:Syz442}
3\,\Jnv{2,2,1}^2 + 3\,\Jnv{3,3,0}\,\Jnv{1,1,2} \textcolor{darkgreen}{+} \Jnv{3,2,1}\,\Jnv{1,2,1} -\Jnv{1,2,1}^2\,\Inv{2,0,0}~=~0\;, \quad\text{and}\quad \mathrm{Y}\leftrightarrow\mathrm{T}\;.
\end{equation}
\begin{equation}\label{eq:Syz341}
3\,\Jnv{2,2,1}\,\Inv{1,2,0} - \Jnv{3,2,1}\,\Inv{0,2,0} + 3\,\Jnv{3,3,0}\,\Inv{0,1,1} \textcolor{darkgreen}{-} \Jnv{1,2,1}\,\Inv{2,2,0}~=~0\;, \quad\text{and}\quad \mathrm{Y}\leftrightarrow\mathrm{T}\;.
\end{equation}
\begin{equation}\label{eq:Syz332I}
3\,\Jnv{2,2,1}\,\Inv{1,1,1} - \Jnv{3,2,1}\,\Inv{0,1,1} + 3\,\Jnv{3,3,0}\,\Inv{0,0,2} \textcolor{darkgreen}{-} \Jnv{1,2,1}\,\Inv{2,1,1}~=~0\;, \quad\text{and}\quad \mathrm{Y}\leftrightarrow\mathrm{T}\;.
\end{equation}
\begin{equation}\label{eq:Syz332II}
3\,\Jnv{2,1,2}\,\Inv{1,2,0} + \Jnv{3,1,2}\,\Inv{0,2,0} - \Jnv{3,2,1}\,\Inv{0,1,1} - \Jnv{1,1,2}\,\Inv{2,0,0}\,\Inv{0,2,0}-\Jnv{1,2,1}\,\Inv{2,0,0}\,\Inv{0,1,1}+\Jnv{1,2,1}\,\Inv{2,1,1}~=~0\;, \quad\text{and}\quad \mathrm{Y}\leftrightarrow\mathrm{T}\;.
\end{equation}
The colored signs here correct sign mistakes in \cite{Trautner:2018ipq} caused by a change of conventions. 
Finally, the syzygy of order [633] needed to derive Eq.~\eqref{eq:633} is given by 
\begin{equation}\label{eq:Syz633}
 \begin{split}
54\,\Jnv{3,3,0}\,\Jnv{3,0,3}~=&~
+90\,\Jnv{2,2,1}\,\Jnv{2,1,2}\,\Inv{2,0,0}
-18\,\Jnv{3,2,1}\,\Jnv{1,1,2}\,\Inv{2,0,0}
-18\,\Jnv{3,1,2}\,\Jnv{1,2,1}\,\Inv{2,0,0}\\
&
-135\,\Jnv{2,2,1}\,\Jnv{1,1,2}\,\Inv{3,0,0}
-135\,\Jnv{2,1,2}\,\Jnv{1,2,1}\,\Inv{3,0,0}\\
&
-54\,\Inv{2,1,1}\,\Inv{1,1,1}\,\Inv{3,0,0}\,\Inv{0,1,1}
+18\,\Inv{1,1,1}^2\,\Inv{2,0,0}^2\,\Inv{0,1,1}
-108\,\Inv{1,1,1}\,\Inv{3,0,0}\,\Inv{1,2,0}\,\Inv{1,0,2}\\
&
+2\,\Inv{2,1,1}\,\Inv{2,0,0}\left(\Inv{2,2,0}\,\Inv{0,0,2}+\Inv{2,0,2}\,\Inv{0,2,0}\right)
+2\,\Inv{2,2,0}\,\Inv{2,0,2}\left(\Inv{2,0,0}\,\Inv{0,1,1}-\Inv{2,1,1}\right)\\
&
-9\,\Inv{3,0,0}\,\Inv{0,1,1}\left(\Inv{2,2,0}\,\Inv{1,0,2}+\Inv{2,0,2}\,\Inv{1,2,0}\right)
-9\,\Inv{3,0,0}\,\Inv{1,1,1}\left(\Inv{2,2,0}\,\Inv{0,0,2}+\Inv{2,0,2}\,\Inv{0,2,0}\right)\\
&
+9\,\Inv{2,0,0}\,\Inv{1,1,1}\left(\Inv{2,2,0}\,\Inv{1,0,2}+\Inv{2,0,2}\,\Inv{1,2,0}\right)
-9\,\Inv{3,0,0}\,\Inv{2,1,1}\left(\Inv{1,2,0}\,\Inv{0,0,2}+\Inv{1,0,2}\,\Inv{0,2,0}\right)\\
&
+8\,\Inv{0,1,1}\,\Inv{0,2,0}\,\Inv{0,0,2}\left(27\,\Inv{3,0,0}^2-\Inv{2,0,0}^3\right)\;.
 \end{split}
\end{equation}

\bibliography{bibliography}

\end{document}